%% file: MassHierarchy_v2010_11_22.tex
\documentclass[12pt]{article}
\usepackage{amssymb,amsmath}

\newcommand{\hu}[1][]{ {h^u_{#1}} }
\newcommand{\hd}[1][]{ {h^d_{#1}} }
\newcommand{\he}[1][]{ {h^e_{#1}} }

\begin{document}
\setcounter{page}{0} \thispagestyle{empty} \vspace*{-0.5cm}
\begin{flushright}
OSU-HEP-10-08
\end{flushright}
\vspace{0.5cm}

\begin{center}
{\Large \textsc{Fermion Mass Hierarchy from\\Symmetry Breaking at the TeV Scale} }

\vspace*{1.5cm}
    B.~N.~Grossmann\footnote{Email address:  benjamin.grossmann@okstate.edu},
    Z.~Murdock\footnote{Email address:  zeke.murdock@okstate.edu}, and
    S.~Nandi\footnote{Email address:  s.nandi@okstate.edu}

\vspace*{0.5cm}
\textit{Department of Physics and Oklahoma Center for High Energy Physics,\\
Oklahoma State University, Stillwater, OK 74078, USA}
\end{center}

\vspace*{0.3in}
\begin{abstract}
In this paper we explore models in which we add a singlet scalar
Higgs to the Standard Model.  All of these models explain the origin
of the mass hierarchy amongst the fermion masses and mixing angles.
We discuss 24 different variations on this model, and explore the
different phenomenological possibilities.  We find that the
phenomenological implications of all the models are very similar
except in the Higgs sector.  Higgs decays and signals can be altered
very significantly in all the models, but break up into two distinct
classes. We also describe a systematic method for generating these
models from higher order interactions involving vector-like quarks
and flavon scalars.
\end{abstract}
\section{Introduction}
One of the major ingredients to the Standard Model (SM) is the mass
generating Higgs mechanism. It works remarkably well for describing
mass relationships among the electroweak bosons.  However, the mass
relationships among the fermions is more obscure because the Yukawa
couplings don't have a predicted structure within the SM.  They span
over five orders of magnitude for no apparent reason.

Many attempts have been made to generate the mass hierarchy of the fermions \cite{attempts}.  Radiative corrections have been used to generate the hierarchy as in \cite{Nandi:2008zw}\cite{Dobrescu:2008sz}\cite{Balakrishna:1987qd}.  Warped extra dimensions were used in \cite{ArkaniHamed:1998rs} and \cite{Gogberashvili:2007gg} where the extra dimensions are creatively ``apple-shaped''.

Previous works \cite{Babu}\cite{Giudice}\cite{Lykken} describe the Yukawa
couplings as dependent on functions of Higgs fields, acting as
higher dimensional operators in the Lagrangian.  The mass hierarchy
of the quarks has a correspondence to the exponents of these
operators. The top quark, having  dimension four  Yukawa
interaction, has an exponent of zero on its operator; thus its
Yukawa coupling is the same as it is in the SM.  For the remaining
quarks, the effective Yukawa interactions are successively higher
and higher dimensional as we include the lighter quarks, and hence
as the exponent increases, the masses get smaller. As an explanation
for the origin of these higher dimensional operators, a
Froggatt-Nielsen \cite{Froggat} type mechanism was used
\cite{Giudice}\cite{Lykken}.

The operators considered previously were composed from either only a
Higgs doublet or only a Higgs singlet.  This paper generalizes this
to allow the possibility for the operators to consist of both a
doublet and a singlet, with the previous works being the limiting
cases.

\section{Effective Model}
\subsection{Modeling the Yukawa Couplings}

The higher dimensional operators in the Yukawa couplings make the
terms non-renormalizable.  However, this is only an effective theory
below a scale $M$, the mass scale where vector-like quarks exist.  The operators in Ref.~\cite{Babu} rely on the SM Higgs doublet $H$, and
are of the form
\begin{align}\label{eqn:operatorH}
\left(\frac{H^\dagger H}{M^2}\right)^{n}
&\hu[ij]\overline{q}_L^iu_R^j\widetilde{H},&
\left(\frac{H^\dagger H}{M^2}\right)^{n} &
\hd[ij] \overline{q}_L^id_R^jH,
\end{align}
where $\widetilde{H}=i\sigma_2H^*$.  The exponent ${n}$ is a
non-negative integer that may have a different value for each pair
of generation indices $i$, $j$. The coupling coefficients $\hu,\hd$
are $\mathcal{O}(1)$. The quark doublet $q_L$ and quark singlets
$u_R$, $d_R$ are the SM quarks. In contrast, the higher dimensional
operators in Ref.~\cite{Lykken} were made by replacing the doublet
operator $H^\dagger H$ with the operator $S^\dagger S$, where $S$
field is a complex scalar singlet under the SM.
\begin{align}\label{eqn:operatorS}
\left(\frac{S^\dagger
S}{M^2}\right)^{n}&\hu[ij]\overline{q}_L^iu_R^j\widetilde{H},&
\left(\frac{S^\dagger
S}{M^2}\right)^{n}&\hd[ij]\overline{q}_L^id_R^jH.
\end{align}

With this setup, the singlet scalar S acts as the messenger of the
EW symmetry breaking for all the quarks and the leptons except the
top quark.

In the unitary gauge, the parametrization is
\begin{align} \label{eqn:vevs}
H &= \begin{pmatrix}0\\ v+\frac{1}{\sqrt{2}}H_0\end{pmatrix},&
S &= \left(v_S+\frac{1}{\sqrt{2}}S_0\right).
\end{align}
with $v\simeq174~\textrm{GeV}$.  The value of $v_S$ is chosen to
also be in the electroweak scale. After assigning vacuum expectation values (vevs) to the Higgs
fields, the resulting operators can be written in terms of
dimensionless parameters.
\begin{align}
\epsilon&=\frac{v}{M},&
\alpha&=\frac{v_S}{v}.
\end{align}

The mass terms resulting from Eq.~\eqref{eqn:operatorH} have a
coefficient $\epsilon^{2n}=(v/M)^{2n}$.
\begin{align}
\epsilon^{2n} &\hu[ij]\overline{q}_L^iu_R^jv,& \epsilon^{2n}
&\hd[ij] \overline{q}_L^id_R^jv.
\end{align}
Similarly, the mass terms resulting from Eq.~\eqref{eqn:operatorS}
have a coefficient $(\alpha\epsilon)^{2n}=(v_S/M)^{2}$.
\begin{align}
(\alpha\epsilon)^{2n} &\hu[ij]\overline{q}_L^iu_R^jv,&
(\alpha\epsilon)^{2n} &\hd[ij] \overline{q}_L^id_R^jv.
\end{align}

Effective dimension four Yukawa couplings, and the mass matrices can
then be constructed by using powers of $\epsilon$ and $\alpha$.
Since $h^u$, $h^d$ are $\mathcal{O}(1)$, the powers needed must
reduce each matrix element by an appropriate order of magnitude
below the top mass. By comparing masses of the top and bottom quarks
with the smallest allowed choice for the exponent $n$, an estimate
on the value of $\epsilon$ can be established.
\begin{align}\label{eqn:epsilonOrder}
\frac{m_b}{m_t}\sim\frac{M^d_{33}}{M^u_{33}}&=\frac{\hd[33]\epsilon^2v}{\hu[33]
v}\sim \epsilon^2 \sim \mathcal{O}(10^{-2}),
\end{align}
and consequently $M$ is around the 1--2 TeV range \cite{Giudice}.

Knowing the experimentally determined quark masses and the CKM
mixing angles, and the order of magnitude of $\epsilon^2$, many
possible mass matrices corresponding to Eq.~\eqref{eqn:operatorH}
can be written down.
The mass matrices for the up and down quarks used in Ref.~\cite{Babu} were

\begin{align}\label{eqn:matEpsilon}
M^u &= \begin{pmatrix}
\hu[11]\epsilon^6  & \hu[12]\epsilon^4    & \hu[13]\epsilon^4    \\
\hu[21]\epsilon^4  & \hu[22]\epsilon^2    & \hu[23]\epsilon^2    \\
\hu[31]\epsilon^4  & \hu[32]\epsilon^2    & \hu[33]
\end{pmatrix}v,&
M^d &= \begin{pmatrix}
\hd[11]\epsilon^6  & \hd[12]\epsilon^6    & \hd[13]\epsilon^6    \\
\hd[21]\epsilon^6  & \hd[22]\epsilon^4    & \hd[23]\epsilon^4    \\
\hd[31]\epsilon^6  & \hd[32]\epsilon^4    & \hd[33]\epsilon^2
\end{pmatrix}v.
\end{align}

Similarly, for the mass matrices from Eq.~\eqref{eqn:operatorS}, as
used in Ref.~\cite{Lykken}, make the replacement
$\epsilon^{2n}\mapsto(\alpha\epsilon)^{2n}$ for each element of
Eq.~\eqref{eqn:matEpsilon}.

These two models take the higher dimensional operators to be purely
composed of only Higgs doublets or only Higgs singlets.  But now
consider the general possibility of operators built from
combinations of doublets and singlets.
\begin{align}\label{eqn:operatorHS}
\left(\frac{H^\dagger H}{M^2}\right)^{n-n_S}\left(\frac{S^\dagger
S}{M^2}\right)^{n_S} &\hu[ij]\overline{q}_L^iu_R^j\widetilde{H},&
\left(\frac{H^\dagger H}{M^2}\right)^{n-n_S}\left(\frac{S^\dagger
S}{M^2}\right)^{n_S} &\hd[ij] \overline{q}_L^id_R^jH.
\end{align}

Since $\epsilon$ regulates the order of magnitude for the mass
matrices, it is reasonable to keep the same $\epsilon$-texture as in
Eq.~\eqref{eqn:matEpsilon}.  This fixes the value of $n$ that will
be used for each matrix element.  However, this does not restrict
the value of $n_S$.  Within the mass matrices $M^u$ and $M^d$, there
are 17 matrix elements that may have varying values for the integer
$n_S$, only restricted by $0\leq n_S \leq n$.

For a mass term with a factor of $\epsilon^{2n}$, the allowed powers
of $\alpha$ are the even numbers ranging from zero to $2n$.
 Inspection of the mass matrices shows there are four mass terms with
$\epsilon^2$ (two allowed powers of $\alpha$), seven with
$\epsilon^4$ (three allowed powers of $\alpha$), and six with
$\epsilon^6$ (four allowed powers of $\alpha$).  Allowing powers of
$\alpha$ to be independent for matrix elements with the same power
of $\epsilon$ means there can be $2^4 3^7 4^6 = 143~327~232$
possible Lagrangians.

To simplify the situation, all matrix elements that have the same
power of $\epsilon$ are restricted to have a common power of
$\alpha$. For example, the mass terms $M^u_{22}$ and $M^u_{23}$ are
both proportional to $\epsilon^2$.  They are restricted to be both
be proportional to $\alpha^0$ or both be proportional to $\alpha^2$.
They are not allow to have different powers of $\alpha$.  This
restriction reduces the possible Lagrangians down to
$2\cdot3\cdot4=24$, a much more manageable number.

With this restriction, the effective Lagrangian can be written down
with common higher dimensional operators factored as leading
coefficients.
\begin{align}\begin{split}
\mathcal{L}^\textrm{Yuk}_{\textrm{quark}}&=\hu[33] \overline{q}_L^{3} u_R^{3} \widetilde{H}\\
    &\quad+\left(\frac{H^\dagger H}{M^2}\right)^{1-n_1} \left(\frac{S^\dagger S}{M^2}\right)^{n_1}
        \left(
          \hd[33] \overline{q}_L^{3} d_R^{3} H
        + \hu[22] \overline{q}_L^{2} u_R^{2} \widetilde{H}
        + \hu[23] \overline{q}_L^{2} u_R^{3} \widetilde{H}
        + \hu[32] \overline{q}_L^{3} u_R^{2} \widetilde{H}
        \right)\\
    &\quad+\left(\frac{H^\dagger H}{M^2}\right)^{2-n_2} \left(\frac{S^\dagger S}{M^2}\right)^{n_2}
        \left(
          \hd[22] \overline{q}_L^{2} d_R^{2} H
        + \hd[23] \overline{q}_L^{2} d_R^{3} H
        + \hd[32] \overline{q}_L^{3} d_R^{2} H
        + \hu[12] \overline{q}_L^{1} u_R^{2} \widetilde{H}
        \right.\\
    &\phantom{\quad+\left(\frac{H^\dagger H}{M^2}\right)^{2-n_2} \left(\frac{S^\dagger S}{M^2}\right)^{n_2}}
        \left.
        + \hu[21] \overline{q}_L^{2} u_R^{1} \widetilde{H}
        + \hu[13] \overline{q}_L^{1} u_R^{3} \widetilde{H}
        + \hu[31] \overline{q}_L^{3} u_R^{1} \widetilde{H}
        \right)\\
    &\quad+\left(\frac{H^\dagger H}{M^2}\right)^{3-n_3}\left(\frac{S^\dagger S}{M^2}\right)^{n_3}
        \left(
          \hd[11] \overline{q}_L^{1}d_R^{1} H
        + \hd[12] \overline{q}_L^{1}d_R^{2} H
        + \hd[21]\overline{q}_L^{2}d_R^{1} H
        \right.\\
    &\phantom{\quad+\left(\frac{H^\dagger H}{M^2}\right)^{3-n_3}\left(\frac{S^\dagger S}{M^2}\right)^{n_3}}
        \left.
        + \hd[13] \overline{q}_L^{1}d_R^{3} H
        + \hd[31] \overline{q}_L^{3}d_R^{1} H
        + \hu[11]\overline{q}_L^{1} u_R^{1} \widetilde{H}
        \right)+h.c..
\end{split}
\end{align}
The operator exponents from Eq.~\eqref{eqn:operatorHS} have been
specified. Since $n$ can takes a fixed value according to the
flavors of each coupling, it can be directly set $(n=1,2,3)$.  Also,
since $0\leq n_S\leq n$, the exponent $n_S$ can be specified as
$n_k$ such that $0\leq n_k\leq k$ where $k\in\{1,2,3\}$.

General expressions for the effective mass matrices and effective
Yukawa couplings in the gauge basis of the quarks can be found in
terms of the of the dimensionless parameters $\alpha$, $\epsilon$
and the exponents $n$, $n_S$.
\begin{align}\label{eqn:generalMasses}
M^u_{ij}&=\alpha^{2n_S}\epsilon^{2n}\hu[ij]v,&
M^d_{ij}&=\alpha^{2n_S}\epsilon^{2n}\hd[ij]v,\\
\label{eqn:h0Yukawa}
f^{hu}_{ij}&=(2(n-n_S)+1)\alpha^{2n_S}\epsilon^{2n}\hu[ij],&
f^{hd}_{ij}&=(2(n-n_S)+1)\alpha^{2n_S}\epsilon^{2n}\hd[ij],\\
\label{eqn:s0Yukawa}
f^{su}_{ij}&=2n_S\alpha^{2n_S-1}\epsilon^{2n}\hu[ij],&
f^{sd}_{ij}&=2n_S\alpha^{2n_S-1}\epsilon^{2n}\hd[ij].
\end{align}

One of the consequences of these higher dimensional operators is
flavor changing neutral currents in the Higgs sector.  This occurs
because the effective Yukawa matrices and the effective mass
matrices will not be proportional. The effective mass matrices will
all be similar to Eq.~\eqref{eqn:matEpsilon} with extra factors of
$\alpha$. However, the effective Yukawa matrices will have extra
numerical factors corresponding to the coefficients resulting from
the binomial expansions of the operators after they acquire vevs. In
Eqs.~\eqref{eqn:generalMasses}--\eqref{eqn:s0Yukawa}, factors with
$n$ or $n_S$ will generally be different for each matrix element.

Generalized expressions for the quark masses can be found in terms
of $\alpha,\epsilon$ and $n_1,n_2,n_3$ using a biunitary
transformation $M^{x}_\textrm{diag}={V^{x}_L}^\dagger M^{x}V^{x}_R$,
where $x\in\{u,d\}$ so that $M^{x}$ is either mass matrix from
Eq.~\eqref{eqn:generalMasses}.

The effective Yukawa matrices  can also be found by $Y^{hx}
={V^x_L}^\dagger f^{hx} V^x_R$, where $V^x_L$, $V^x_R$ are the
unitary matrices that diagonalized $M^x$. Since the Yukawa matrices
$f^{hx}$ ,$f^{sx}$ are not proportional to the mass matrices $M^x$,
the Yukawa matrices in the mass basis $Y^{hx}$, $Y^{sx}$ will not be
diagonal.

Expansions of the mass and Yukawa matrices were made in powers of
$\epsilon$.  The masses and Yukawa couplings have been expanded to
$\epsilon^6$.  The CKM matrix is also given up to $\epsilon^4$. The
Calculations were made assuming coupling coefficients are real and
symmetric ($h^x_{ij}=h^x_{ji}$). The masses are listed below.  The
Yukawa couplings and CKM matrix are in the Appendix.
\begin{subequations}
\begin{align}
{M^u_\textrm{diag}}_{11}&\approx
\left(\alpha^{2n_3}\hu[11]-\alpha^{2(2n_2-n_1)} \frac{ \hu[12]^2 }{ \hu[22] }\right)~\epsilon^6,\\
{M^u_\textrm{diag}}_{22}&\approx
\alpha^{2n_1} \hu[22]~\epsilon^2
-\alpha^{4n_1} \frac{ \hu[23]^2 }{ \hu[33] }~\epsilon^4
+\left(\alpha^{2(2n_2-n_1)}\frac{ \hu[12]^2 }{ \hu[22] } +\alpha^{6n_1} \frac{\hu[22]\hu[23]^2 }{\hu[33]^2}\right)~\epsilon^6,\\
{M^u_\textrm{diag}}_{33}&\approx
\hu[33]
+ \alpha^{4n_1} \frac{\hu[23]^2}{\hu[33]}~\epsilon^4
+ \alpha^{6n_1} \frac{\hu[22]\hu[23]^2}{\hu[33]^2}~\epsilon^6,\\
{M^d_\textrm{diag}}_{11}&\approx
\alpha^{2n_3}\hd[11]~\epsilon^6,\\
{M^d_\textrm{diag}}_{22}&\approx
\alpha^{2n_2}\hd[22]~\epsilon^4
-\alpha^{2(2n_2-n_1)} \frac{\hd[23]^2}{\hd[33]}~\epsilon^6,\\
{M^d_\textrm{diag}}_{33}&\approx \alpha^{2n_1}\hd[33]~\epsilon^2
+\left(\alpha^{2(2n_2-n_1)}\frac{ \hd[23]^2}{\hd[23]}
-\alpha^{2(6n_2-3n_1-2n_3)}
\frac{\hd[22]^2\hd[23]^4}{\hd[13]^2\hd[33]^3} \right)~\epsilon^6.
\end{align}
\end{subequations}

The Yukawa couplings $f^{hx},f^{sx}$ are couplings of the quarks to
the Higgs fields with both field types in the gauge basis. The
Yukawa couplings $Y^{hx},Y^{sx}$ are with the quarks in the mass
basis, however, the Higgs fields are still in the gauge basis. To
get the Yukawa couplings in the mass basis of both the quarks and
the Higgs fields, the rotation of the Higgs fields still needs to be
applied (see Sec~\ref{sec:higgs}).  Doing so yields couplings $Y^{'hx},Y^{'sx}$ of the form
\begin{subequations}
\begin{align}
\frac{1}{\sqrt{2}}Y^{'hx}_{ij}&=\frac{1}{\sqrt{2}}\left(Y^{hx}_{ij}\cos\beta-Y^{sx}_{ij}\sin\beta\right)&\textrm{Coupling to Higgs mass eigenstate }h,\\
\frac{1}{\sqrt{2}}Y^{'sx}_{ij}&=\frac{1}{\sqrt{2}}\left(Y^{hx}_{ij}\sin\beta+Y^{sx}_{ij}\cos\beta\right)&\textrm{Coupling to Higgs mass eigenstate }s.
\end{align}
\end{subequations}

\subsection{Higgs Sector and $Z^\prime$}\label{sec:higgs}
In the effective Lagrangian, the Higgs boson $S$ only appears in the
product $S^\dag{}S$.  This means it is free to be charged under an
additional gauge symmetry while all the SM fields are neutral under
this new symmetry.  This symmetry is assumed to be a $U(1)_S$ local
symmetry.  It plays a role---along with extra flavor symmetries---in
the mechanism for creating the effective Lagrangian.  This will be
expanded upon in Section~\ref{sec:mechanism}.

The general Higgs potential which mixes the SM Higgs doublet $H$
with the $S$ is
\begin{align}
V(H,S)&=-\mu_H^2 H^\dagger H -\mu_S^2 S^\dagger S
    +\lambda_H(H^\dagger H)^2
    +\lambda_{HS}(H^\dagger H)(S^\dagger S)
    +\lambda_S(S^\dagger S)^2.
\end{align}

Minimization of the potential to find in terms of the vevs yields
\begin{align}
\mu_H^2&=v^2(2\lambda_H+\alpha^2\lambda_{HS}),&
\mu_S^2&=v^2(2\alpha^2\lambda_{S}+\lambda_{HS}).
\end{align}
The squared mass matrix in the $(H_0,S_0)$
basis is
\begin{align}
\mathcal{M}^2&=2v^2\begin{pmatrix}
2\lambda_{H}&\alpha\lambda_{HS}\\
\alpha\lambda_{HS}&2\alpha^2\lambda_{S}
\end{pmatrix}.
\end{align}
The mass eigenstate basis $(h,s)$ can be written using
\begin{align}\label{eqn:mix}
H_0 &= h \cos\theta + s \sin\theta,&
S_0 &= - h \sin\theta + s\cos\theta,
\end{align}
where the Higgs sector mixing angle $\theta$ is expressible with
\begin{align}
\tan2\theta&=\frac{\alpha\lambda_{HS}}{\lambda_H-\alpha^2\lambda_{S}}.
\end{align}
The physical squared masses at the tree level are
\begin{align}
m_{h,s}^2&=2v^2\left(\lambda_H+\alpha^2\lambda_S\mp\sqrt{(\lambda_H-\alpha^2\lambda_S)^2+\alpha^2\lambda_{HS}^2}\right).
\end{align}

The $Z^\prime$ gauge boson gets its mass from the pseudoscalar
component of $S$ when the $U(1)_S$ symmetry is broken.  Assuming the
$S$ has a charge of 1 under this symmetry, and the coupling constant
is $g_S$, then
\begin{align}
m_{Z^\prime}^2&=2 g_S^2 v_S^2.
\end{align}

The vev $v_S$ is of the order of the EW scale.  If the coupling
$g_S\sim\mathcal{O}(1)$, then the mass of the $Z^\prime$ should also
be expected to be near the EW scale.  However, since $g_S$ isn't
determined, the mass may be different.  There is no significant bound on the mass of the $Z^\prime$ from LEP \cite{lepewwg}.

Since the $Z^\prime$ does not couple to SM fields directly, its
presence can only be determined from interactions with new fields
and mixing with the $Z$.  Mixing can occur through kinetic mixing or
from higher order loop effects.  Measurements of the $Z$ properties
at LEP1 constrain the mixing to be $\lesssim10^{-3}$
\cite{Yao}\cite{Langacker}, for $m_{Z^\prime}< 1$ TeV.

The $Z^\prime$ will couple to the SM fermions via dimension 6 interactions.  No significant bound is placed on $M_{Z^\prime}$ from these interactions as was shown in \cite{dob}

\section{Phenomenology}
There are 24 different models that can be composed under the current
scheme.  Covering the specific phenomenology of each model would not
be very illuminating.  The differences can be seen by looking at two
different types of signals that can be seen at colliders:  Higgs
decay signatures and FCNC processes.
\subsection{Higgs Decays}
In the low Higgs mass range of the SM, allowed by the LEP limit, say
114--130 GeV, the dominant mode of Higgs decay is to two bottom
quarks $h\rightarrow b \overline{b}$.  The branching ratio for this
decay is almost 100\%. This is undesirable from an experimental
point of view because these signals are difficult to disentangle
from a large QCD background. In all of the 24 models, the branching
ratio of this decay can be altered significantly. The 24 models
break up into two distinct classes. These two classes are
characterized by whether we have $(H^{\dag} H/M^2)$ or $(S^{\dag}
S/M^2)$ in the prefactor in the 2nd line of Eq.~(10).

For the 12 models coming from $S^{\dag}S$ in Eq. (10), $h\rightarrow
b {\bar b}$ coupling is
\begin{align}
\hd[33] (\alpha \cos\theta - 2 \sin\theta)\epsilon^2
\alpha/\sqrt{2}.
\end{align}

Taking $\alpha \sim 1$ and $\theta \sim 26^\circ$, this coupling is
reduced significantly compared to that in the SM making the
observation of the $h \rightarrow \gamma \gamma$ signal at the LHC
much more favorable compared to the SM. For example, the
$h\rightarrow\gamma\gamma$ signal can be enhanced by a factor of 10
as seen in Ref.~\cite{Lykken}\cite{tevatron_search}\cite{Melni}. Varying the angle $\theta$
drastically alters the branching ratio of the Higgs. If there is no
mixing, $\theta \sim 0^\circ$, the structure of Higgs decays is
virtually indistinguishable from the Standard Model.

In the other 12 models, with the prefactor in Eq. (10), line 2 given
by $H^{\dag} H /M^2$, the coupling of $h\rightarrow b \overline{b}$ is $3 \epsilon^2 \hd[33] \cos\theta$. This causes an enhancement of $9 \cos^2
\theta$ for the $h\rightarrow b \overline{b}$.

The rate for the $h\rightarrow\gamma\gamma$ mode will only change by
a factor of $\cos^2\theta$. Since $h\rightarrow b \overline{b}$ is
enhanced by a factor of 9, while the rate for the
$h\rightarrow\gamma\gamma$ mode stays the same, the branching ratio
for the $\gamma\gamma$ mode is effectively reduced by a factor of 9.
Thus in this class of models, the observation of the Higgs signal in
the $\gamma \gamma$ mode will be much more difficult than in the SM.
\subsection{FCNC Processes}
\subsubsection{$t\rightarrow ch$}

The observation of the decay mode $t\rightarrow ch$ will be a clear
indication for physics beyond the SM. For the SM, this branching
ratio is very tiny $\sim10^{-14}$ \cite{tchsm}. In the models considered here, this
branching ratio can be much larger, and observable at the LHC.  For
the 12 $H^{\dag}H$ models, the coupling for the $t\rightarrow ch$
mode is $\hu[32]\sqrt{2}\epsilon^2 \cos\theta$, giving rise to a
branching ratio $\sim10^{-4}$, with $\hu[23]\sim1$ and $\cos\theta\sim1$.
With large top quark production at the LHC, this decay
mode with such a branching ratio will be observable at the LHC. For
the other 12 models with $S^{\dag}S$, the coupling for $t\rightarrow
ch$ mode is $\hu[32]\sqrt{2}\epsilon^2 \sin\theta$. This BR is also much larger than in the SM, however
much smaller than the models with $H^{\dag}H$. Furthermore, there
is possibilty of cancellation leading to further reduction in the
BR.

\subsubsection{$B_s^0 \rightarrow \mu^+\mu^-$}
In all 24 models this process is mediated by $s$ and $h$ exchange.
Amongst the 24 models there are 3 different categories of amplitudes
for this process.  The couplings for the 6 different categories of
models are controlled by $(H^\dagger H)^2$, $(S^\dagger S)^2$, and
$(H^\dagger H)( S^\dagger S)$.

\begin{subequations}
\begin{align}
( H^\dagger H)& - (H^\dagger H)^2:
 &A_{h} & \sim  -5\epsilon^8 \he[22]\hd[23] \cos^2\theta / m_h^2\\
&&A_{s} & \sim  -5\epsilon^8 \he[22]\he[23] \sin^2\theta / m_s^2\\
( H^\dagger H)&-(H^\dagger H)( S^\dagger S):
 &A_{h} & \sim  \alpha^2\epsilon^8 \he[22]\hd[23] \sin\theta (3\alpha \cos\theta - 2 \sin\theta)/m_h^2\\
&&A_{s} & \sim  -\alpha^2\epsilon^8 \he[22]\hd[23] \cos\theta(3\alpha \sin\theta + 2 \cos\theta)/m_s^2\\
( H^\dagger H)& -( S^\dagger S)^2:
 &A_{h} & \sim  \alpha^6\epsilon^8 \he[22]\hd[23] ( \alpha^2 \cos^2\theta - 8\sin^2\theta-\alpha \sin 2\theta)/m_h^2\\
&&A_{s} & \sim  \alpha^6\epsilon^8 \he[22]\hd[23] ( \alpha^2\sin^2\theta - 8\cos^2\theta+\alpha \sin 2\theta)/m_s^2\\
( S^\dagger S)& - (H^\dagger H)^2:
&A_{h} & \sim  -5\epsilon^8 \he[22]\hd[23] \cos\theta (2 \alpha \cos\theta +  \sin\theta)/\alpha m_h^2\\
&&A_{s} & \sim  -5\epsilon^8 \he[22]\hd[23] \sin\theta (2 \alpha\sin\theta -  \cos\theta)/\alpha m_s^2\\
( S^\dagger S)& -(H^\dagger H)( S^\dagger S):
&A_{h} & \sim  \alpha^3\epsilon^8 \he[22]\hd[23] \cos\theta (2 \sin\theta - 3\alpha \cos\theta)/m_h^2\\
&&A_{s} & \sim  -\alpha^3\epsilon^8 \he[22]\hd[23] \sin\theta (2\cos\theta + 3\alpha \sin\theta)/m_s^2\\
( S^\dagger S)& - ( S^\dagger S)^2:
&A_{h} & \sim  \alpha^6\epsilon^8 \he[22]\hd[23] \sin \theta (\alpha \cos\theta - 4 \sin\theta)/m_h^2\\
&&A_{s} & \sim  -\alpha^6\epsilon^8 \he[22]\hd[23] \cos \theta (\alpha\sin\theta + 4\cos\theta)/m_s^2
\end{align}
\end{subequations}
While there are differences in the form of each amplitude, they are
all proportional to $\epsilon^8$.  This means the
$\textrm{BR}\sim10^{-9}$ is still within $< 4.7\times10^{-8}$, the
experimental limit \cite{Lykken}.
\subsection{Double Higgs Production}

It is possible to pair produce the vector-like quarks of our model.
The dominant decay mode (~95\%) is to a SM quark and a Higgs (e.g.
$Q_L \rightarrow u h$).  The Higgs will then decay to $b \bar{b}$.
The signal for this process will be 4 $b$-jets and 2 hard jets.
Taking the mass of the vector-like quarks to be 1 TeV, the
production cross section for pair production at the LHC at 14 TeV is
\~60 fb per each new quark \cite{cacc}.  Because there are more than 50 new quarks
in each model, the total production cross section for all new quarks
will be $\sim30$ pb. We place kinematic cuts on the signal and
background as follows:  the invariant mass of the $b$-jets,
$m_{bb}>100\ \mathrm{GeV}$, the $p^{\textrm{jets}}_T > 100 \
\mathrm{GeV}$ and for the $b$-jets, $p^{b}_T > 30 \ \mathrm{GeV}$.
Using CalcHEP we find that imposing these cuts will reduce the
branching ratio of the new quarks to $\textrm{BR}(Q_L \rightarrow u
h) \sim 0.9$ fb. If we take the $b$-tagging efficiency to be $50\%$,
the signal is reduced by a factor of $\frac{1}{16}$ .  With these
cuts and $10\ \textrm{fb}^{-1}$ of luminosity we would expect to see
$\sim 30$ events per additional quark in the model.

The SM background for a 6 $b$ final state was calculated in
Ref.~\cite{Grossmann:2010wm}.  With their cuts the background is 60
fb. With a $100 \ \mathrm{GeV}$ cut on each of the final state
non-$b$-jets, we expect that the background for $bbbbjj$ in the SM
will be of similar order.  With a few extra vector-like quarks
from one of the models the signal should be much larger than the
background and observable at the LHC with enough luminosity.

\section{Mass Generating Mechanism}\label{sec:mechanism}
So far, the gauge symmetries of the SM have been extended by an
additional $U(1)_S$ local symmetry.  To explain the origin of the
operators ${H^\dagger}H$ and ${S^\dagger}S$ in the Yukawa couplings,
some additional $U(1)_{F_i}$ global symmetries will also be employed
in the use of a Froggatt-Nielsen type mechanism.  In the one model
where none of the effective low energy interactions in the
Lagrangian have a coefficient of $({S^\dagger}S)^{n_S}$, the
$U(1)_S$ symmetry is not included; it essentially decouples from the
model, so can be ignored.

Each $U(1)_{F_i}$ global symmetry has a flavor scalar boson $F_i$
(called a flavon) that is charged only under it's corresponding
symmetry, and neutral with all other symmetries.  Because the
$U(1)_{F_i}$ are global symmetries, there are no gauge bosons
associated with them. It should be noted that even though there is
no restriction being placed on the number of $U(1)_{F_i}$
symmetries, it is not necessary to have more than two.

The effective Yukawa couplings are created by interactions with new
heavy exotic quarks, the new flavons $F_i$, and the Higgs bosons $H$
and $S$. These quarks will be denoted by $Q$ for the doublets, $U$
and $D$ for the singlets. They have the same hypercharges as their
SM counterparts $q$, $u$, and $d$.

These extra fields are necessary in a Froggatt-Nielsen mechanism as
each field occupies a unique position in the charge space of
$U(1)_Y\times{}U(1)_S\times{}U(1)_{F_1}\times{}U(1)_{F_2}\cdots$.
The flavons and the Higgs fields provide the interactions that link
the fields to their neighbors in the charge space.  A sequence of
interactions is required to move between non-neighboring fields,
such as the SM quarks.


Within a given model, the sequence of field interactions beginning
and ending with SM particles was chosen so that there is only one
sequence connecting any two SM quarks (assuming backward steps
aren't taken within the sequence). Although this isn't strictly
necessary, if distinctly different sequences of particle
interactions were allowed into a model, then some models may have
explicit terms in the Lagrangian with higher powers of
$({H^\dagger}H)^{n-n_S}({S^\dagger}S)^{n_S}$ than are written down.

In order to make different interaction sequences  non-interacting
with each other, it is necessary to space the non-interacting quark
fields at least two quantum numbers away from each other in the
charge space.   This leads to a large number of quark fields being
used as each interaction sequence path through this space must be
long enough to go around many other fields and avoid interacting
with other paths.

It should be noted that unlike the SM quarks, which only has right
handed singlets and left handed doublets, the new heavy quarks occur
in left-right pairs and behave vector-like with respect to the gauge
groups of the SM and $U(1)_S$.  The quantum numbers of a left-right
pair will be identical except for the quantum number of one
$U(1)_{F_i}$ symmetry.  This quantum number will differ by a value
of one.  When this symmetry breaks, the vev of the $F_i$ gives mass
to the new heavy quarks.  The vev of each $F_i$ is are assumed to be
around the TeV scale.

\subsection{Couplings in the Lagrangian}
The couplings of these heavy quarks in the Lagrangian take on a
generalized form.  Specifically, the couplings each particle has
will depend on their charge assignments within a given model.  The
hermitian conjugate of each coupling will also be included in the
Lagrangian.  For the terms listed in
eqs.~\eqref{eqn:coupling1}--\eqref{eqn:coupling3}, no summation over
the indices is implied.
\begin{align}
\left.\begin{aligned}
&f^{FQ}_{ab}\overline{Q_L^a}Q_R^bF_{i}           &&f^{FU}_{ab}\overline{U_L^a}U_R^bF_{i}            &&f^{FD}_{ab}\overline{D_L^a}D_R^bF_{i}\\
&f^{FQ}_{ab}\overline{Q_L^a}Q_R^bF_{i}^\dag
&&f^{FU}_{ab}\overline{U_L^a}U_R^bF_{i}^\dag
&&f^{FD}_{ab}\overline{D_L^a}D_R^bF_{i}^\dag
\end{aligned}\right\}&\textrm{Flavon Couplings}\label{eqn:coupling1}\\
\left.\begin{aligned}
&f^{SQ}_{ab}\overline{Q_L^a}Q_R^bS        &&f^{SU}_{ab}\overline{U_L^a}U_R^bS         &&f^{SD}_{ab}\overline{D_L^a}D_R^bS\\
&f^{SQ}_{ab}\overline{Q_L^a}Q_R^bS^\dag
&&f^{SU}_{ab}\overline{U_L^a}U_R^bS^\dag
&&f^{SD}_{ab}\overline{D_L^a}D_R^bS^\dag
\end{aligned}\right\}&\textrm{Higgs Singlet Couplings}\label{eqn:coupling2}\\
\left.
\begin{aligned}
&f^{HU}_{ab}\overline{Q_L^a}U_R^b\widetilde{H}   &&f^{HD}_{ab}\overline{Q_L^a}D_R^bH\\
&f^{HU}_{ab}\overline{Q_R^a}U_L^b\widetilde{H}
&&f^{HD}_{ab}\overline{Q_R^a}D_L^bH
\end{aligned}\right\}&\textrm{Higgs Doublet Couplings}\label{eqn:coupling3}
\end{align}
All of the coupling coefficients
($f^{FQ},f^{FU},f^{FD},f^{SQ},f^{SU},f^{SD},f^{HU},f^{HD}$) are
taken to be $\mathcal{O}(1)$.

Every heavy quark will have one coupling from
Eq.~\eqref{eqn:coupling1} where $a=b$, as this will be a massive
left-right pair when the flavon $F$ breaks the flavor symmetry.
Every heavy quark must also have at least one coupling from
eqs.~\eqref{eqn:coupling1}--\eqref{eqn:coupling2} where $a\neq b$,
or from Eq.~\eqref{eqn:coupling3} where $a$ and $b$ are indexed over
different quark types. For the few heavy quarks that directly couple
to the SM quarks, the appropriate replacement should be made to the
terms coming from eqs.~\eqref{eqn:coupling1}--\eqref{eqn:coupling3}
(e.g. \(\overline{Q_L}Q_R\mapsto \overline{q_L}Q_R\)).
\subsubsection{The Effective Lagrangian}
In all model variations, the Yukawa coupling
\(\hu[33]\overline{q_L^3}u_R^3\widetilde{H}\) is the only one that
involves only SM particles.  All the other model variations have
coefficients of \((H^\dag H)^{n-n_S}(S^\dag S)^{n_S}\) on terms which
would otherwise be SM Yukawa couplings. These terms are larger than
four dimensions and come from a process of integrating out the heavy
fermions from the tree level diagrams, which correspond with terms
of the forms from eqs.~\eqref{eqn:coupling1}--\eqref{eqn:coupling3}.

\begin{figure}[t]
\begin{center}
\begin{picture}(308,40)
\put(0,15){\line(1,0){308}}
\multiput(22,15)(22,0){13}{\line(0,1){2}}
\multiput(22,19)(22,0){13}{\line(0,1){2}}
\multiput(22,23)(22,0){13}{\line(0,1){2}}
\multiput(22,27)(22,0){13}{\vector(0,1){2}}
\multiput(22,31)(22,0){13}{\line(0,1){2}}
\multiput(22,35)(22,0){13}{\line(0,1){2}}
\multiput(11,15)(22,0){14}{\vector(1,0){5}}
\put(20,40){\(\widetilde{H}\)} \multiput(42,40)(22,0){3}{\(F_{1}\)}
\put(108,40){\(H\)} \multiput(130,40)(22,0){3}{\(F_{1}\)}
\put(196,40){\(\widetilde{H}\)}
\multiput(218,40)(22,0){4}{\(F_{1}\)} \put(6,0){\(q_L^2\)}
\put(28,0){\(U_R^4\)} \put(50,0){\(U_L^4\)} \put(72,0){\(U_R^5\)}
\put(94,0){\(U_L^5\)} \put(116,0){\(Q_R^6\)} \put(138,0){\(Q_L^6\)}
\put(160,0){\(Q_R^7\)} \put(182,0){\(Q_L^7\)} \put(204,0){\(U_R^8\)}
\put(226,0){\(U_L^8\)} \put(248,0){\(U_R^9\)} \put(270,0){\(U_L^9\)}
\put(292,0){\(u_R^3\)}
\end{picture}
\caption{Feynman diagram linking $q_L^2$ to $u_R^3$ in an
${H^\dagger}H$ model.}\label{fig:feynman:HH}
\end{center}
\end{figure}
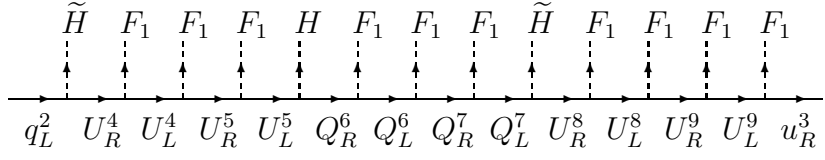

For example, consider the term
\((H^\dag{}H/M^2)\hu[23]\overline{q_L^2}u_R^3\widetilde{H}\). This
terms exists in 12 of the 24 variations of the effective Lagrangian.
One possible heavy quark model has thirteen terms associated with
this process. This process can be represented by the Feynman diagram
in Fig.~\ref{fig:feynman:HH}. The necessary terms are
\begin{align}\label{eqn:Frog1}
\begin{split}
&f^{HU}_{2,4}\overline{q_L^2}U_R^4\widetilde{H}
+f^{FU}_{4,4}\overline{U_R^4}U_L^4F_1
+f^{FU}_{4,5}\overline{U_L^4}U_R^5F_1
+f^{FU}_{5,5}\overline{U_R^5}U_L^5F_1
+f^{HU}_{5,6}\overline{U_L^5}Q_R^6H \\
&+f^{FQ}_{6,6}\overline{Q_R^6}Q_L^6F_1
+f^{FQ}_{6,7}\overline{Q_L^6}Q_R^7F_1
+f^{FQ}_{7,7}\overline{Q_R^7}Q_L^7F_1
+f^{HU}_{7,8}\overline{Q_L^7}U_R^8\widetilde{H}\\
&+f^{FU}_{8,8}\overline{U_R^8}U_L^8F_1
+f^{FU}_{8,9}\overline{U_L^8}U_R^9F_1
+f^{FU}_{9,9}\overline{U_R^9}U_L^9F_1
+f^{FU}_{9,3}\overline{U_L^9}u_R^3F_1+ h.c..
\end{split}
\end{align}

This particular model choice exhibits only a single extra symmetry
\(U(1)_{F_1}\) in the given terms.  The flavon symmetry breaks, and
the flavons acquire a vev $\langle F_i\rangle$.  The heavy fermions
can be integrated out and an effective expression belos the TeV
scale is proportional to
\begin{align}
f^{HU}_{2,4} f^{FU}_{4,4} f^{FU}_{4,5} f^{FU}_{5,5} f^{HU}_{5,6}
f^{FQ}_{6,6} f^{FQ}_{6,7} f^{FQ}_{7,7} f^{HU}_{7,8} f^{FU}_{8,8}
f^{FU}_{8,9} f^{FU}_{9,9} f^{FU}_{9,3}
\left(\frac{\langle{F_1}\rangle}{M}\right)^{10} \frac{H^\dag H}{M^2}
\overline{q_L^2} u_R^3\widetilde{H}+ h.c..
\end{align}
Thus for this particular model choice, the effective coupling
parameter in the low energy Lagrangian is
\begin{align}
\hu[23]\sim f^{HU}_{2,4} f^{FU}_{4,4} f^{FU}_{4,5} f^{FU}_{5,5}
f^{HU}_{5,6} f^{FQ}_{6,6} f^{FQ}_{6,7} f^{FQ}_{7,7} f^{HU}_{7,8}
f^{FU}_{8,8} f^{FU}_{8,9} f^{FU}_{9,9} f^{FU}_{9,3}
\left(\frac{\langle{F_1}\rangle}{M}\right)^{10}.
\end{align}
The couplings $f$ are $\mathcal{O}(1)$.  The vev of the flavons is
the same order as the vector-like quark masses,
$\langle{F_1}\rangle\sim M$.  This means $\hu$ can also be
$\mathcal{O}(1)$, consistent with the assumption in the effective
Lagrangian.

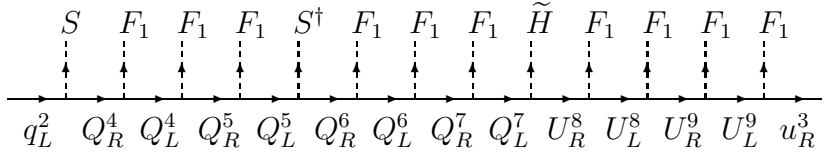
\begin{figure}[t]
\begin{center}
\begin{picture}(308,40)
\put(0,15){\line(1,0){308}}
\multiput(22,15)(22,0){13}{\line(0,1){2}}
\multiput(22,19)(22,0){13}{\line(0,1){2}}
\multiput(22,23)(22,0){13}{\line(0,1){2}}
\multiput(22,27)(22,0){13}{\vector(0,1){2}}
\multiput(22,31)(22,0){13}{\line(0,1){2}}
\multiput(22,35)(22,0){13}{\line(0,1){2}}
\multiput(11,15)(22,0){14}{\vector(1,0){5}} \put(20,40){\(S\)}
\multiput(42,40)(22,0){3}{\(F_{1}\)} \put(108,40){\(S^\dag\)}
\multiput(130,40)(22,0){3}{\(F_{1}\)}
\put(196,40){\(\widetilde{H}\)}
\multiput(218,40)(22,0){4}{\(F_{1}\)} \put(6,0){\(q_L^2\)}
\put(28,0){\(Q_R^4\)} \put(50,0){\(Q_L^4\)} \put(72,0){\(Q_R^5\)}
\put(94,0){\(Q_L^5\)} \put(116,0){\(Q_R^6\)} \put(138,0){\(Q_L^6\)}
\put(160,0){\(Q_R^7\)} \put(182,0){\(Q_L^7\)} \put(204,0){\(U_R^8\)}
\put(226,0){\(U_L^8\)} \put(248,0){\(U_R^9\)} \put(270,0){\(U_L^9\)}
\put(292,0){\(u_R^3\)}
\end{picture}
\caption{Feynman diagram linking $q_L^2$ to $u_R^3$ in an
${S^\dagger}S$ model.}\label{fig:feynman:SS}
\end{center}
\end{figure}

For a comparison, consider the 12 effective Lagrangians that have
the same term, except with $(H^\dag{}H)\mapsto(S^\dag{}S)$. Some
possible model choices may again have thirteen terms.  The Feynman
diagram in Fig.~\ref{fig:feynman:SS} is similar to Fig.~\ref{fig:feynman:HH}, but there are noticeable differences in the
first 5 interactions.
\begin{align}\label{eqn:Frog2}
\begin{split}
&f^{SQ}_{2,4}\overline{q_L^2}Q_R^4S
+f^{FQ}_{4,4}\overline{Q_R^4}Q_L^4F_1
+f^{FQ}_{4,5}\overline{Q_L^4}Q_R^5F_1
+f^{FQ}_{5,5}\overline{Q_R^5}Q_L^5F_1
+f^{SQ}_{5,6}\overline{Q_L^5}Q_R^6S^\dag\\
&+f^{FQ}_{6,5}\overline{Q_R^6}Q_L^6F_1
+f^{FQ}_{6,7}\overline{Q_L^6}Q_R^7F_1
+f^{FQ}_{7,7}\overline{Q_R^7}Q_L^7F_1
+f^{HU}_{7,8}\overline{Q_L^7}U_R^8\widetilde{H}\\
&+f^{FU}_{8,8}\overline{U_R^8}U_L^8F_1
+f^{FU}_{8,9}\overline{U_L^8}U_R^9F_1
+f^{FU}_{9,9}\overline{U_R^9}U_L^9F_1
+f^{FU}_{9,3}\overline{U_L^9}u_R^3F_1+ h.c..
\end{split}
\end{align}

This expression is similar to the previous case, and likewise when
the heavy fermions are integrated out, the effective expression
below the TeV scale looks similar.
\begin{align}
f^{SQ}_{2,4} f^{FQ}_{4,4} f^{FQ}_{4,5} f^{FQ}_{5,5} f^{SQ}_{5,6}
f^{FQ}_{6,5} f^{FQ}_{6,7} f^{FQ}_{7,7} f^{HU}_{7,8} f^{FU}_{8,8}
f^{FU}_{8,9} f^{FU}_{9,9} f^{FU}_{9,3}
\left(\frac{\langle{F_1}\rangle}{M}\right)^{10} \frac{S^\dag S}{M^2}
\overline{q_L^2} u_R^3\widetilde{H}+ h.c..
\end{align}
As before, the effective coupling parameters in the low energy
Lagrangian can be pulled out
\begin{align}
\hu[23]\sim f^{SQ}_{2,4} f^{FQ}_{4,4} f^{FQ}_{4,5} f^{FQ}_{5,5}
f^{SQ}_{5,6} f^{FQ}_{6,5} f^{FQ}_{6,7} f^{FQ}_{7,7} f^{HU}_{7,8}
f^{FU}_{8,8} f^{FU}_{8,9} f^{FU}_{9,9} f^{FU}_{9,3}
\left(\frac{\langle{F_1}\rangle}{M}\right)^{10},
\end{align}
and once again, $\hu[23]\sim\mathcal{O}(1)$.

\subsection{Charge Assignments for Specific Models}
\subsubsection{Effective Lagrangian with only powers of $(S^\dag S/M^2)$}
In the effective Lagrangian where all the interactions of dimension
greater than four only have powers of $(S^\dag S/M^2)$, the $U(1)_S$
symmetry only needs to be accompanied with only one $U(1)_F$
symmetry.  With only these two extra symmetries, one possible model
has a set of 166 heavy quarks (18 $U_L^a$, 18 $U_R^a$, 29 $D_L^a$,
29 $D_R^a$, 36 $Q_L^a$, 36 $Q_R^a$).  The quantum numbers of the SM
quarks, Higgs doublet, Higgs singlet, and the vector boson are in
Table \ref{tbl:StypeSM}. The quantum numbers of the heavy quarks can
be found in Table \ref{tbl:StypeQUD}.
\begin{table}[t]
\caption{Charge assignments of the SM quarks, Higgs doublet, Higgs
singlet, and new vector boson for an effective Lagrangian with only
powers of $(S^\dag S/M^2)$.}\label{tbl:StypeSM}
\begin{center}
\begin{tabular}{|c|c|c||c|c|c||c|c|c|}\hline
Fields   &   $U(1)_S$  &   $U(1)_F$  &   Fields   &   $U(1)_S$ &
$U(1)_F$&   Fields   &   $U(1)_S$ &   $U(1)_F$\\\hline
$q_L^3$  &   0   &   0   &$u_R^3$  &   0   &   0   &   $d_R^3$  &   0   &   4   \\
$q_L^2$  &   0   &   16  &$u_R^2$  &   0   &   4   &   $d_R^2$  &   0   &   10  \\
$q_L^1$  &   0   &   24  &$u_R^1$  &   0   &   10  &   $d_R^1$  & 0
&   32  \\\hline $H$      &   0   &   0   &   $F$      &   0   & 1
\\\cline{4-6} $S$      &   1   &   0   \\\cline{1-3}
\end{tabular}
\end{center}
\end{table}
\subsubsection{Effective Lagrangian with only powers of $(H^\dag H/M^2)$}
In the effective Lagrangian where there are no interactions with the
$S$, the $U(1)_S$ symmetry is effectively eliminated.

Some possible choices of fields for a model use only two
$U(1)_{F_i}$ symmetries.  As indicated previously, this means there
will be two new bosons $F_1$ and $F_2$.  And explicit choice can be
made that consists of 124 heavy quarks (17 $U_L^a$, 17 $U_R^a$, 20
$D_L^a$, 20 $D_R^a$, 20 $Q_L^a$, 20 $Q_R^a$).  The quantum numbers
of the SM quarks, Higgs doublet, and the vector bosons are in
Table~\ref{tbl:HtypeSM}.  The heavy quarks have their quantum
numbers listed in Table \ref{tbl:HtypeQUD}.
\begin{table}[t]
\caption{Charge assignments of the SM quarks, Higgs doublet, and new
vector bosons, for an effective Lagrangian with only powers of
$(H^\dag H/M^2)$.}\label{tbl:HtypeSM}
\begin{center}
\begin{tabular}{|c|c|c||c|c|c||c|c|c|}\hline
Fields   &   $U(1)_{F_1}$  &   $U(1)_{F_2}$  &   Fields   &
$U(1)_{F_1}$  &   $U(1)_{F_2}$&   Fields   & $U(1)_{F_1}$  &
$U(1)_{F_2}$\\\hline
$q_L^3$  &   0   &   0   &$u_R^3$  &   0   &   0   &   $d_R^3$  &   5   &   -5  \\
$q_L^2$  &   -2  &   0   &$u_R^2$  &   2   &   6   &    $d_R^2$  &   5   &   3   \\
$q_L^1$  &   -4  &   -2  &$u_R^1$  &   5   &   3   &   $d_R^1$  & -6
&   4   \\\hline $H$      &   0   &   0   &$F_1$    &   1   & 0
\\\cline{1-3} \multicolumn{3}{c|}{}    &$F_2$    &   0   &   1
\\\cline{4-6}
\end{tabular}
\end{center}
\end{table}

\subsubsection{Generalized Model}
The two previous models were constructed independently from each
other.  Each of the other 22 models can also be constructed
independently from each other.  However, constructing a model and
assigning appropriate charges to the heavy quarks for each effective
Lagrangian doesn't need to be done 24 times.  It is possible to
construct a generalized model that will match any of the effective
Lagrangians by simply changing specific groups of particles.

An example of this change was already done in
Eqs.~\eqref{eqn:Frog1} and \eqref{eqn:Frog2}.  In those
expressions, the fields \(U_R^4\), \(U_L^4\), \(U_R^5\), and
\(U_L^5\) were replaced with \(Q_R^4\), \(Q_L^4\), \(Q_R^5\), and
\(Q_L^5\); and the corresponding interactions with \(H\) were
replaced to interactions with \(S\). In terms of their quantum
numbers, the change in hypercharge of the \(U(1)_Y\) symmetry at
either end of the sequence was replaced for a change in the charge
of the \(U(1)_S\) symmetry.

The generalized model presented here uses two \(U(1)_{F_i}\)
symmetries and has 282 heavy quarks.  The charge assignments of the
fields under these symmetries can be found in Tables
\ref{tbl:GeneralSM}, and \ref{tbl:GeneralQ}--\ref{tbl:GeneralUD}.

\begin{table}[t]
\caption{Charge assignments of the SM quarks, Higgs doublet, Higgs
singlet, and the new vector bosons, for a generalized
model.}\label{tbl:GeneralSM}
\begin{center}
\begin{tabular}{|c|c|c|c||c|c|c|c|}\hline
Fields   &   \(U(1)_S\)  &   \(U(1)_{F_1}\)  &   \(U(1)_{F_2}\)  &
Fields   &   \(U(1)_S\)  &   \(U(1)_{F_1}\)  &
\(U(1)_{F_2}\)\\\hline
\(q_L^3\)  &   0   &   5   &   12  &   \(d_R^3\)  &   0   &   7   &   18  \\
\(q_L^2\)  &   0   &   5   &   2   &   \(d_R^2\)  &   0   &   13  &   20  \\
\(q_L^1\)  &   0   &   6   &   29  &   \(d_R^1\)  &   0   &   0   &
17  \\\hline
\(u_R^3\)  &   0   &   5   &   12  &   \(H\)       &   0   &   0   &   0   \\
\(u_R^2\)  &   0   &   11  &   14  &   \(S\)       &   1   &   0   &
0   \\\cline{5-8} \(u_R^1\)  &   0   &   13  &   20  &    \(F_1\) &
0   &   1   &   0   \\\cline{1-4} \multicolumn{3}{c}{}       & &
\(F_2\)     &   0   &   0   &   1   \\\cline{5-8}
\end{tabular}
\end{center}
\end{table}

As presented, Tables \ref{tbl:GeneralQ} and \ref{tbl:GeneralUD} are
for the effective Lagrangian with only powers of \((H^\dag H/M^2)\).
To adjust the table to fit any of the other variations of the
Lagrangian, replace an appropriate set of fields with a different
set of fields. The choice of replacements is also, in most cases,
not unique.  A particular choice of replacements is given in Tables
\ref{tbl:Replacements1}--\ref{tbl:Replacements6}.

It should be noted, in the replacement tables the numbering
subscript is changed to avoid possible duplication of names.  For
example, the replacement \(U_L^{20}\mapsto Q_L^{90}\) is made
instead of \(U_L^{20}\mapsto Q_L^{20}\) because there already exists
a quark with the name \(Q_L^{20}\).
\subsection{Charged Leptons}
So far, the mass hierarchy of the quarks has been addressed, but the
hierarchy of the leptons hasn't been mentioned.  Fortunately, the
same approach of using vector-like leptons can be used.  A mass
matrix like those from Eqn.~\eqref{eqn:matEpsilon} can be
constructed for the charged leptons.  It turns out, the matrix $M^e$
can have the same $\epsilon$ texture as $M^d$.

Thus generally, the mass matrix has and the Yukawa couplings have
the form
\begin{align}
M^e_{ij}&=\alpha^{2n_S}\epsilon^{2n}\he[ij]\frac{v_H}{\sqrt{2}},\\
f^{he}_{ij}&=(2(n-n_S)+1)\alpha^{2n_S}\epsilon^{2n}\he[ij],\\
f^{se}_{ij}&=2n_S\alpha^{2n_S-1}\epsilon^{2n}\he[ij].
\end{align}
Again, the values of $n$ and $n_S$ may vary between matrix elements

Because it has the same powers of $\epsilon$ as the down-type quark
sector, it is not always necessary to find a set of vector-like
leptons from scratch.  If a set of vector-like quarks is known, then
some simple replacements can be made.
\begin{align}
D&\mapsto E, &Q&\mapsto L,    &U&\mapsto N.
\end{align}
Then remove the vector-like leptons that are unnecessary.  This will
eliminate the unwanted interactions with the light neutrinos.nteractions with the light neutrinos.

\section{Conclusions}

Presented here is a scheme under which the fermion mass hierarchy
can be understood by couplings with other massive vector-like
fermions.  The effective Yukawa couplings are generated by the
breaking of global flavor symmetries $U(1)_{F_i}$ at the TeV scale.
It should be noted, as an effective model, it may not be valid above
the breaking scale.  At that point, another mechanism may take over,
or the theory may be embedded in a larger symmetry group.

The variations of the effective model have decays and exchange
amplitudes that are different, based upon the interactions with
Higgs doublets $H$ and singlets $S$.  Phenomena that distinguishes
between variations of the effective model have Branching Ratios that
should fall within the observable limits of the LHC.

\section*{Acknowledgments}
We thank J.~D.~Lykken for useful discussions.  This work is supported in part by the United States Department of Energy, Grant Numbers DE-FG02-04ER41306 and DE-FG02-04ER46140.

\include{MassHierarchy_tables}


\end{document}

%% file: MassHierarchy_tables.tex
\section{Appendix}
\subsection{Yukawa Couplings in the Quark Mass Basis}

The expansions made in this section use the unitary transformation matrices that diagonalize the mass matrices, $M_{\textrm{diag}}=V_L^\dagger M V_R$.   It is assumed the mass matrices are symmetric.  Consequently, the unitary transformation matrices are equal, $V_L=V_R$.

The $\overline{u}uH_0$-Yukawa couplings in the $u$-mass eigenbasis are
$Y^{hu}={V^u_L}^\dagger f^{hu} V^u_R$.
\begin{subequations}
\begin{flalign}
Y^{hu}_{11} &\approx\left((4 n_2-2n_1-7)\alpha^{2( n_2-n_1)}\frac{{\hu[12]}^2}{\hu[22]}+(7-2 n_3)\alpha^{2n_3}\hu[11]\right)~\epsilon^6&\\
\begin{split}
Y^{hu}_{12} &\approx 2 (n_2 - n_1- 1) \alpha^{2 n_2} \hu[12]~\epsilon^4+2 \alpha^{2(n_1 + n_2)}\hu[23] \left((2-n_2)\frac{\hu[13]}{\hu[33]} + (n_1-1)\frac{\hu[12] \hu[23]}{\hu[22] \hu[33]}\right)~\epsilon^6
\end{split}&\\
\begin{split}
Y^{hu}_{13} &\approx
{2\alpha^{2n_2}}\left((2 - n_2) \hu[13] + (n_1-1)\frac{ \hu[12] \hu[23]}{\hu[22]}\right)~\epsilon^4\\
&\quad+2\alpha^{2(n_1+n_2)}\hu[23] \left( (1+n_1-n_2)\frac{\hu[12]}{\hu[33]}+(1-n_1)\frac{\hu[13]\hu[23]}{\hu[22]\hu[33]}+(n_1-1)\frac{\hu[12]{\hu[23]}^2}{{\hu[22]}^2\hu[33]} \right)~\epsilon^6
\end{split}&\\
\begin{split}
Y^{hu}_{22} &\approx
(3 - 2 n_1) \alpha^{2 n_1}\hu[22] ~\epsilon^2+(4 n_1-5)\alpha^{4 n_1}   \frac{{\hu[23]}^2}{\hu[33]}~\epsilon^4\\
&\quad+ \left((6n_1-7)  \alpha^{6 n_1}\frac{{\hu[22]}
{\hu[23]}^2}{{\hu[33]}^2} + (7 + 2 n_1 - 4 n_2) \alpha^{2( n_2-n_1)}\frac{{\hu[12]}^2 }{\hu[22]}\right)~\epsilon^6
\end{split}&\\
\begin{split}
Y^{hu}_{23} &\approx
2( n_1-1)\alpha^{2 n_1} \hu[23] ~\epsilon^2+2 (n_1-1)\alpha^{4 n_1}  \frac{\hu[22] \hu[23]}{\hu[33]}~\epsilon^4\\
&\quad+\left(2(n_1-1)\alpha^{6 n_1}\hu[23]\frac{{\hu[22]}^2-2{\hu[23]}^2}{{\hu[33]}^2}\right.\\
&\quad\qquad\left.+2(n_2-2)\alpha^{2(n_2-n_1)}\frac{\hu[12]\hu[13]}{\hu[22]}+(1-n_1)\alpha^{4n_2-2n_1}\frac{{\hu[12]}^2\hu[23]}{{\hu[22]}^2}\right)
~\epsilon^6
\end{split}&\\
\begin{split}
Y^{hu}_{33} &\approx
\hu[33]+(4 n_1-5)\alpha^{4 n_1}\frac{{\hu[23]}^2}{\hu[33]}~\epsilon^4
+(7 - 6 n_1) \alpha^{6 n_1}\frac{\hu[22] {\hu[23]}^2 }{{\hu[33]}^2}~\epsilon^6
\end{split}&
\end{flalign}
\end{subequations}

The $\overline{u}uS_0$-Yukawa couplings in the $u$-mass eigenbasis are
$Y^{su}={V^u_L}^\dagger f^{su} V^u_R$.
\begin{subequations}
\begin{flalign}
\begin{split}
Y^{su}_{11} &\approx
2\left((n_1-2n_2)\alpha^{4 n_2-2n_1-1}\frac{{\hu[12]}^2}{\hu[22]}+n_3\alpha^{2n_3-1}\hu[11]\right)~\epsilon^6
\end{split}&\\
\begin{split}
Y^{su}_{12} &\approx
2 (n_1-n_2) \alpha^{2 n_2-1} \hu[12]~\epsilon^4
+2 \alpha^{2(n_1 + n_2)-1}\hu[23]\frac{n_2\hu[13]\hu[22]-n_1\hu[12]\hu[23]}{\hu[22]\hu[33]}~\epsilon^6
\end{split}&\\
\begin{split}
Y^{su}_{13} &\approx
2\alpha^{2n_2-1}\frac{n_2 \hu[13] \hu[22]-n_1 \hu[12] \hu[23]}{\hu[22]}~\epsilon^4\\
&\quad+2\alpha^{2(n_1+n_2)-1}\hu[23] \left(
(n_2-n_1)\frac{\hu[12]}{\hu[33]}+n_1\frac{\hu[13]\hu[23]}{\hu[22]\hu[33]}-n_1\frac{\hu[12]{\hu[23]}^2}{{\hu[22]}^2\hu[33]}
\right)~\epsilon^6
\end{split}&\\
\begin{split}
Y^{su}_{22} &\approx
2 n_1 \alpha^{2 n_1-1}\hu[22]~\epsilon^2
-4 n_1 \alpha^{4 n_1-1}\frac{{\hu[23]}^2}{\hu[33]}~\epsilon^4\\
&\quad+
 \left(
- 6n_1  \alpha^{6 n_1-1}\frac{\hu[22]
{\hu[23]}^2}{{\hu[33]}^2} - 2(n_1 - 2 n_2) \alpha^{4 n_2-2n_1-1}\frac{{\hu[12]}^2 }{\hu[22]}
\right)~\epsilon^6
\end{split}&\\
\begin{split}
Y^{su}_{23} &\approx
-2n_1\alpha^{2 n_1-1} \hu[23]~\epsilon^2
-2 n_1\alpha^{4 n_1-1}  \frac{\hu[22] \hu[23]}{\hu[33]}~\epsilon^4\\
&\quad+
\left(2n_1\alpha^{6 n_1-1}\hu[23]\frac{2{\hu[23]}^2-{\hu[22]}^2}{{\hu[33]}^2}+\alpha^{4n_2-2n_1-1}\hu[12]\frac{n_1\hu[12]\hu[23]-2n_2\hu[13]\hu[22]}{{\hu[22]}^2}\right)
~\epsilon^6
\end{split}&\\
\begin{split}
Y^{su}_{33} &\approx
4 n_1\alpha^{4n_1-1}\frac{{\hu[23]}^2}{\hu[33]}~\epsilon^4+
6 n_1\alpha^{6n_1-1}\frac{\hu[22]{\hu[23]}^2}{{\hu[33]}^2}~\epsilon^6
\end{split}&
\end{flalign}
\end{subequations}

The $\overline{d}dH_0$-Yukawa couplings in the $d$-mass eigenbasis are
$Y^{hd}={V^d_L}^\dagger f^{hd} V^d_R$.
\begin{subequations}
\begin{flalign}
Y^{hd}_{11}&\approx
(7-2n_3)\alpha^{2n_3}\hd[11]~\epsilon^6&\\
Y^{hd}_{12}&\approx
2(n_3-n_2-1)\alpha^{2n_3}\hd[12]~\epsilon^6&\\
Y^{hd}_{13}&\approx
2\alpha^{2n_3}\left(
(4+2n_1-2n_3)\hd[13]+2(n_2-n_1-1)\frac{\hd[12]\hd[23]}{\hd[22]}\right)~\epsilon^6&\\
Y^{hd}_{22}&\approx
(5-2n_2)\alpha^{2n_2}\hd[22]~\epsilon^4
+(4n_2-2n_1-7)\alpha^{2(n_2-n_1)}\frac{{\hd[23]}^2}{\hd[33]}~\epsilon^6&\\
Y^{hd}_{23}&\approx
2(n_2-n_1-1)\alpha^{2n_2}\hd[23]~\epsilon^4
+2(n_2-n_1-1)\alpha^{2(n_2-n_1)}\frac{\hd[22]\hd[23]}{\hd[33]}~\epsilon^6&\\
\begin{split}
Y^{hd}_{33}&\approx
(3-2n_1)\alpha^{2 n_1} \hd[33]~\epsilon^2\\
&\quad+\left(
(7+2n_1-4n_2)\alpha^{2(n_2-n_1)}\frac{{\hd[23]}^2}{\hd[33]}+
(2n_1-3)\alpha^{2(6n_2-3n_1-2n_3)}\frac{{\hd[22]}^2{\hd[23]}^4}{{\hd[13]}^2}{{\hd[33]}^3}
\right)~\epsilon^6
\end{split}&
\end{flalign}
\end{subequations}

The $\overline{d}dS_0$-Yukawa couplings in the $d$-mass eigenbasis are
$Y^{sd}={V^d_L}^\dagger f^{sd} V^d_R$.
\begin{subequations}
\begin{flalign}
Y^{sd}_{11}&\approx
2n_3\alpha^{2n_3-1}\hd[11]~\epsilon^6&\\
Y^{sd}_{12}&\approx
2(n_2-n_3)\alpha^{2n_3-1}\hd[12]~\epsilon^6&\\
Y^{sd}_{13}&\approx
2\alpha^{2n_3-1}\left((n_3-n_1)\hd[13]+(n_1-n_2)\frac{\hd[12]\hd[23]}{\hd[22]}\right)~\epsilon^6&\\
Y^{sd}_{22}&\approx
2n_2\alpha^{2n_2-1}\hd[22]~\epsilon^4
+2(n_1-2n_2)\alpha^{4n_2-2n_1-1}\frac{{\hd[23]}^2}{\hd[33]}~\epsilon^6&\\
Y^{sd}_{23}&\approx
2(n_1-n_2)\alpha^{2n_2-1}\hd[23]~\epsilon^4
+2(n_1-n_2)\alpha^{4n_2-2n_1-1}\frac{\hd[22]\hd[23]}{\hd[33]}~\epsilon^6&\\
\begin{split}
Y^{sd}_{33}&\approx
2n_1\alpha^{2 n_1-1} \hd[33]~\epsilon^2\\
&\quad+2\left((2n_2-n_1)\alpha^{4n_2-2n_1-1}\frac{{\hd[23]}^2}{\hd[33]}
-n_1\alpha^{12n_2-6n_1-4n_3-1}\frac{{\hd[22]}^2{\hd[23]}^4}{{\hd[13]}^2{\hd[33]}^3}
\right)~\epsilon^6
\end{split}&
\end{flalign}
\end{subequations}

\subsection{CKM Matrix}
The Cabbio-Kobayashi-Maskawa matrix is
$V^\textrm{CKM}={V^u_L}^\dagger V^d_L$.
\begin{subequations}
\begin{flalign}
\begin{split}
V^{\textrm{CKM}}_{11}&\approx   1-
\left( \frac{\alpha^{4(n_2-n_1)}}{2}\frac{{\hu[12]}^2}{{\hu[22]}^2}+\frac{\alpha^{4(n_3-n_2)}}{2}\frac{{\hd[12]}^2}{{\hd[22]}^2}-\alpha^{2(n_3-n_1)}\frac{\hd[12]\hu[12]}{\hd[22]\hu[22]}\right)\epsilon^4
\end{split}&
\end{flalign}
\begin{flalign}
\begin{split}
V^{\textrm{CKM}}_{21}&\approx   \left(\alpha^{2(n_3-n_2)}\frac{\hd[12]}{\hd[22]}-\alpha^{2(n_2-n_1)}\frac{\hu[12]}{\hu[22]}\right)~\epsilon^2\\
&\quad+\left(
\alpha^{2(4n_1-n_2)} {\hu[23]}^2 \frac{ {\hu[23]}^2 - {\hu[22]}^2 }{ \hu[12] {\hu[33]}^2 }
-\alpha^{2n_2}  \frac{ \hu[13] \hu[23] }{ \hu[22] \hu[33] }\right.\\
&\quad\qquad\left.+\alpha^{2(n_3 - n_1)} \hd[23] \frac{ \hd[12] \hd[23]-\hd[13] \hd[22] }{ {\hd[22]}^2 \hd[33] }
+\alpha^{4(n_3 - n_2)} \frac{ \hd[11] \hd[12] }{ {\hd[22]}^2 }
\right)~\epsilon^4
\end{split}&
\end{flalign}
\begin{flalign}
\begin{split}
V^{\textrm{CKM}}_{31}&\approx
\left(
\alpha^{4(n_1-n_2)} {\hu[23]}^2 \frac{ {\hu[23]}^2 - {\hu[22]}^2 }{ \hu[13] {\hu[33]}^2 }
+\alpha^{2(n_3-n_1)} \frac{ \hd[12] \hd[23] - \hd[13] \hd[22] } { \hd[22] \hd[23] }\right.\\
&\quad\qquad\left.-\alpha^{2(n_3-n_2+n_1)} \frac{ \hd[12] \hu[23] }{ \hd[22] \hu[33] }
\right)~\epsilon^4
\end{split}&
\end{flalign}
\begin{flalign}
\begin{split}
V^{\textrm{CKM}}_{12}&\approx   \left(\alpha^{2(n_2-n_1)} \frac{\hu[12]}{\hu[22]}-\alpha^{2(3n_2-2n_1-n_3)}\frac{\hd[22]{\hd[23]}^2}{\hd[12]{\hd[33]}^2}\right)~\epsilon^2\\
&\quad+\left(\alpha^{2n_2} \hu[23] \frac{\hu[12]\hu[23]-\hu[13]\hu[22]}{\hu[22]^2\hu[33]}\right.\\
&\quad\qquad\left.+\alpha^{2(4n_2-3n_1-n_3)} \hu[23]^2 \frac{ \hd[12](\hd[23]^2-\hd[22]^2) -\hd[13]\hd[22]\hd[23]  }{ \hd[12]^2\hd[33]^3}
\right)~\epsilon^4
\end{split}&
\end{flalign}
\begin{flalign}
\begin{split}
V^{\textrm{CKM}}_{22}&\approx   1 -
\left(
\alpha^{4n_1}       \frac{ \hu[23]^2 }{2\hu[33]^2}
-\alpha^{2n_2}      \frac{ \hd[23]\hu[23] }{ \hd[33]\hu[33] }
+\alpha^{4(n_2-n_1)}\frac{ \hd[33]^2\hu[12]^2+\hd[23]^2\hu[22]^2 }{ 2\hd[33]^2\hu[22]^2 }\right.\\
&\quad\qquad\left.+\alpha^{4(3n_2-2n_1-n_3)}  \frac{ \hd[22]^2\hd[23]^4 }{ 2\hd[12]^2\hd[33]^4 }
-\alpha^{2(4n_2-3n_1-n_3)}  \frac{ \hd[22]\hd[23]^2\hu[12] }{ \hd[12]\hd[33]^2\hu[22] }
\right)~\epsilon^4
\end{split}&
\end{flalign}
\begin{flalign}
V^{\textrm{CKM}}_{32}&\approx   \left(\alpha^{2(n_2-n_1)} \frac{\hd[23]}{\hd[33]}-\alpha^{2n_1}\frac{\hu[23]}{\hu[33]}\right)~\epsilon^2
+\left(
\alpha^{4(n_2-n_1)} \frac{ \hd[22]\hd[23] }{ \hd[33]^2 }
-\alpha^{4n_1}      \frac{ \hu[22]\hu[23] }{ \hu[33]^2 }
\right)~\epsilon^4&
\end{flalign}
\begin{flalign}
\begin{split}
V^{\textrm{CKM}}_{13}&\approx   \alpha^{2(3n_2-2n_1-n_3)} \frac{\hd[22]{\hd[23]}^2}{\hd[13]{\hd[33]}^2}~\epsilon^2\\
&\quad+\left(
\alpha^{2n_2} \frac{ \hu[12]\hu[23]-\hu[13]\hu[22] }{ \hu[22]\hu[33] }
-\alpha^{4(n_2-n_1)} \frac{ \hd[23]\hu[12] }{\hd[33]\hu[22] }\right.\\
&\quad\qquad\left.
-\alpha^{2(4n_2-3n_1-n_3)} \hd[12]^2 \frac{ \hd[13](\hd[22]^2-\hd[23]^2)-\hd[12]\hd[22]\hd[23] }{\hd[13]^2\hd[33]^3 }
\right)~\epsilon^4
\end{split}&
\end{flalign}
\begin{flalign}
\begin{split}
V^{\textrm{CKM}}_{23}&\approx   \left(\alpha^{2n_1} \frac{\hu[23]}{\hu[33]}-\alpha^{2(n_2-n_1)}\frac{\hd[23]}{\hd[33]}\right)~\epsilon^2\\
&\quad+\left(
\alpha^{4n_1} \frac{ \hu[22]\hu[23] }{ \hu[33]^2 }
-\alpha^{4(n_2-n_1)} \frac{ \hd[22]\hd[23] }{ \hd[33]^2 }
+\alpha^{2(4n_2-3n_1-n_3)} \frac{ \hd[22]\hd[23]^2\hu[12] }{ \hd[13]\hd[33]^2\hu[22] }
\right)~\epsilon^4
\end{split}&
\end{flalign}
\begin{flalign}
\begin{split}
V^{\textrm{CKM}}_{33}&\approx   1 -\left(
\alpha^{4n_1} \frac{ \hu[23]^2 }{ 2\hu[33]^2 }
-\alpha^{2n_2}  \frac{\hd[23]\hu[23] }{ \hd[33]\hu[33] }
+\alpha^{4(n_2-n_1)} \frac{ \hd[23]^2 }{ 2\hd[33]^2 }
+\alpha^{4(3n_2-4n_1-n_3)} \frac{ \hd[22]^2\hd[23]^4 }{ 2\hd[13]^2\hd[33]^4 }
\right)~\epsilon^4
\end{split}&
\end{flalign}
\end{subequations}

\newpage
\subsection{Tables of Charge Assignments for Exotic Quarks}
\begin{table}[!h]
\begin{center}
\caption{Charge assignments of the heavy $Q$ quarks for a model
having an effective Lagrangian with only powers of $(S^\dagger
S/M^2)$.}\label{tbl:StypeQUD}
\begin{tabular}{|c|c|c||c|c|c||c|c|c|}\hline
Fields   &   $U(1)_S$&   $U(1)_F$  &   Fields   &$U(1)_S$
&$U(1)_F$&   Fields   &$U(1)_S$  &$U(1)_F$   \\\hline
$Q^{1}_{L,R}$  &   3   &   29, 28  &$Q^{13}_{L,R}$ &  1   &   23, 24  &$Q^{25}_{L,R}$ &  -2  &   4, 3  \\
$Q^{2}_{L,R}$  &   2   &   6, 7    &$Q^{14}_{L,R}$ &  1   &   27, 26  &$Q^{26}_{L,R}$ &  -2  &   18, 17\\
$Q^{3}_{L,R}$  &   2   &   16, 17  &$Q^{15}_{L,R}$ &  0   &   4, 3    &$Q^{27}_{L,R}$ &  -2  &   20, 19\\
$Q^{4}_{L,R}$  &   2   &   18, 19  &$Q^{16}_{L,R}$ &  0   &   26, 25  &$Q^{28}_{L,R}$ &  -2  &   24, 23\\
$Q^{5}_{L,R}$  &   2   &   20, 21  &$Q^{17}_{L,R}$ &  0   &   30, 29  &$Q^{29}_{L,R}$ &  -2  &   26, 25\\
$Q^{6}_{L,R}$  &   2   &   22, 23  &$Q^{18}_{L,R}$ &  0   &   32, 31  &$Q^{30}_{L,R}$ &  -2  &   28, 27\\
$Q^{7}_{L,R}$  &   2   &   28, 27  &$Q^{19}_{L,R}$ &  -1  &   1, 0    &$Q^{31}_{L,R}$ &  -3  &   5, 4  \\
$Q^{8}_{L,R}$  &   1   &   7, 8    &$Q^{20}_{L,R}$ &  -1  &   3, 2    &$Q^{32}_{L,R}$ &  -3  &   7, 6  \\
$Q^{9}_{L,R}$  &   1   &   9, 10   &$Q^{21}_{L,R}$ &  -1  &   17, 16  &$Q^{33}_{L,R}$ &  -3  &   9, 8  \\
$Q^{10}_{L,R}$ &   1   &   11, 12  &$Q^{22}_{L,R}$ &  -1  &   21, 20  &$Q^{34}_{L,R}$ &  -3  &   11, 10\\
$Q^{11}_{L,R}$ &   1   &   13, 14  &$Q^{23}_{L,R}$ &  -1  &   23, 22  &$Q^{35}_{L,R}$ &  -3  &   13, 12\\
$Q^{12}_{L,R}$ &   1   &   15, 16  &$Q^{24}_{L,R}$ &  -1  &   29, 28  &$Q^{36}_{L,R}$ &  -3  &   15, 14\\\hline
\end{tabular}
\end{center}
\end{table}

\begin{table}
\begin{center}
\caption{Charge assignments of the heavy $U$ and $D$ quarks for a
model having an effective Lagrangian with only powers of $(S^\dagger
S/M^2)$.}\label{tbl:StypeQUD2}
\begin{tabular}{|c|c|c||c|c|c||c|c|c|}\hline
Fields   &   $U(1)_S$&   $U(1)_F$  &   Fields   &$U(1)_S$
&$U(1)_F$&   Fields   &$U(1)_S$  &$U(1)_F$   \\\hline
$U^{1}_{L,R}$  &   1   &   1, 0    &$U^{7}_{L,R}$  &  0   &   8, 9    &$U^{13}_{L,R}$ &-1  &   11, 10\\
$U^{2}_{L,R}$  &   1   &   3, 2    &$U^{8}_{L,R}$  &  0   &   16, 15  &$U^{14}_{L,R}$ &-1  &   15, 14\\
$U^{3}_{L,R}$  &   1   &   5, 4    &$U^{9}_{L,R}$  &  0   &   18, 17  &$U^{15}_{L,R}$ &-2  &   4, 5  \\
$U^{4}_{L,R}$  &   1   &   7, 6    &$U^{10}_{L,R}$ &  0   &   22, 21  &$U^{16}_{L,R}$ &-2  &   6, 7  \\
$U^{5}_{L,R}$  &   1   &   19, 18  &$U^{11}_{L,R}$ &  0   &   24, 23  &$U^{17}_{L,R}$ &-2  &   12, 11\\
$U^{6}_{L,R}$  &   1   &   21, 20  &$U^{12}_{L,R}$ &  -1  &   7, 8    &$U^{18}_{L,R}$ &-2  &   14, 13\\\hline
$D^{1}_{L,R}$  &   3   &   29, 30  &$D^{11}_{L,R}$ &  0   &   22, 21  &$D^{21}_{L,R}$ &   -2  &   30, 31 \\
$D^{2}_{L,R}$  &   2   &   6, 5    &$D^{12}_{L,R}$ &  0   &   24, 23  &$D^{22}_{L,R}$ &   -3  &   15, 16 \\
$D^{3}_{L,R}$  &   2   &   30, 31  &$D^{13}_{L,R}$ &  -1  &   7, 8    &$D^{23}_{L,R}$ &   -3  &   17, 18\\
$D^{4}_{L,R}$  &   1   &   5, 4    &$D^{14}_{L,R}$ &  -1  &   11, 10  &$D^{24}_{L,R}$ & -3&   19, 20\\
$D^{5}_{L,R}$  &   1   &   19, 18  &$D^{15}_{L,R}$ &  -1  &   15, 14  &$D^{25}_{L,R}$ &   -3  &   21, 22\\
$D^{6}_{L,R}$  &   1   &   21, 20  &$D^{16}_{L,R}$ &  -1  &   31, 32  &$D^{26}_{L,R}$ &   -3  &   23, 24\\
$D^{7}_{L,R}$  &   1   &   31, 32  &$D^{17}_{L,R}$ &  -2  &   4, 5    &$D^{27}_{L,R}$ &   -3  &   25, 26 \\
$D^{8}_{L,R}$  &  0   &   8, 9    &$D^{18}_{L,R}$ &  -2  &   6, 7    &$D^{28}_{L,R}$ &   -3  &   27, 28 \\
$D^{9}_{L,R}$  &  0   &   16, 15  &$D^{19}_{L,R}$ &  -2  &   12, 11  &$D^{29}_{L,R}$ & -3 &   29, 30\\\cline{7-9}
$D^{10}_{L,R}$ &  0   &   18, 17  &$D^{20}_{L,R}$ &  -2  &   14, 13  \\\cline{1-6}
\end{tabular}
\end{center}
\end{table}

\begin{table}
\begin{center}
\caption{Charge assignments of the heavy $Q$, $U$, and $D$ quarks
for a model having an effective Lagrangian  with only powers of
$(H^\dag H/M^2)$.} \label{tbl:HtypeQUD}
\begin{tabular}{|c|c|c||c|c|c||c|c|c|}\hline
Fields   &   $U(1)_{F_1}$  &   $U(1)_{F_2}$  &   Fields
&$U(1)_{F_1}$  & $U(1)_{F_2}$&Fields   &   $U(1)_{F_1}$  &
$U(1)_{F_2}$\\\hline
$Q^{1}_{L,R}$   & 5, 4    &   -5, -5  &$Q^{8}_{L,R}$  & 1, 0    &   -3, -3  &$Q^{15}_{L,R}$ & -3, -4  &   -1, -1  \\
$Q^{2}_{L,R}$   & 5, 4    &   3, 3    &$Q^{9}_{L,R}$  & 0, 1    &   4, 4    &$Q^{16}_{L,R}$ & -3, -3  &   1, 0    \\
$Q^{3}_{L,R}$   & 3, 2    &   -5, -5  &$Q^{10}_{L,R}$ & 0, -1   &   6, 6    &$Q^{17}_{L,R}$ & -4, -3  &   6, 6    \\
$Q^{4}_{L,R}$   & 3, 3    &   -1, 0   &$Q^{11}_{L,R}$ & -1, -1  &   -1, -2  &$Q^{18}_{L,R}$ & -4, -4  &   -4, -3  \\
$Q^{5}_{L,R}$   & 3, 3    &   3, 4    &$Q^{12}_{L,R}$ & -1, -2  &   -5, -5  &$Q^{19}_{L,R}$ & -5, -5  &   1, 0    \\
$Q^{6}_{L,R}$   & 2, 1    &   6, 6    &$Q^{13}_{L,R}$ & -1, -1  &   1, 0&$Q^{20}_{L,R}$ & -6, -5  &   4, 4    \\\cline{7-9}
$Q^{7}_{L,R}$   & 1, 1  &   1, 0    &$Q^{14}_{L,R}$ & -2, -3  &   4, 4    \\\hline
$U^{1}_{L,R}$   &  2, 3   &   -1, -1  &$U^{7}_{L,R}$   & -1, -1  &   2, 1  &$U^{13}_{L,R}$ & -3, -3  &   4, 3   \\
$U^{2}_{L,R}$   &  0, 1   &   -1, -1  &$U^{8}_{L,R}$  & -1, -1  &   4, 3   &$U^{14}_{L,R}$ & -5, -4  &   -2, -2 \\
$U^{3}_{L,R}$   &  1, 1   &   2, 1    &$U^{9}_{L,R}$  & -1, -1  &   6, 5   &$U^{15}_{L,R}$ & -5, -4  &   6, 6   \\
$U^{4}_{L,R}$   &  1, 1   &   4, 3    &$U^{10}_{L,R}$ & -2, -2  &   -3, -2 &$U^{16}_{L,R}$ & -5, -5  &   0, -1  \\
$U^{5}_{L,R}$   & -1, -1  &   -4, -5  &$U^{11}_{L,R}$ & -2, -2  &   -1, 0   &$U^{17}_{L,R}$ & -5, -5  &   4, 5   \\\cline{7-9}
$U^{6}_{L,R}$   &  0,-1  &   -3, -3     &$U^{12}_{L,R}$ & -3, -3  &   2, 1   \\\hline
$D^{1}_{L,R}$  &  3, 3   &    0, 1    &$D^{8}_{L,R}$ &  0, 0   &   -1, 0  &$D^{15}_{L,R}$& -3, -2  &    6, 6 \\
$D^{2}_{L,R}$  &  3, 3   &    4, 5    &$D^{9}_{L,R}$ &  1, 0   &    2, 0  &$D^{16}_{L,R}$& -2, -3  &   -5, -5\\
$D^{3}_{L,R}$  &  3, 2   &    2, 2    &$D^{10}_{L,R}$& -1, 0   &    4, 4  &$D^{17}_{L,R}$& -4, -4  &   -5, -4\\
$D^{4}_{L,R}$  &  3, 2   &    6, 6    &$D^{11}_{L,R}$& -1, -2  &   -2, -2 &$D^{18}_{L,R}$& -3, -4  &   -2, -2\\
$D^{5}_{L,R}$  &  2, 1   &   -5, -5   &$D^{12}_{L,R}$& -2, -2  &    1, 0  &$D^{19}_{L,R}$& -3, -4  &    2, 2\\
$D^{6}_{L,R}$  &  1, 1   &   -4, -3   &$D^{13}_{L,R}$& -1, -2  &    2, 2    &$D^{20}_{L,R}$& -5, -5  &    2, 1\\\cline{7-9}
$D^{7}_{L,R}$  &  1, 1   &-2, -1   &$D^{14}_{L,R}$& -2, -2  &    5, 4\\\cline{1-6}
\end{tabular}
\end{center}
\end{table}

\begin{table}
\begin{center}
\caption{Charge assignments for the heavy quark doublets to
be used in a generalized model.  The given values are for the case
where all coefficients are of the form $(H^\dag H/M^2)^n$. See
Tables \ref{tbl:Replacements1}--\ref{tbl:Replacements6} to make the
necessary changes for the different
Lagrangians.}\label{tbl:GeneralQ}
\begin{tabular}{|c|c|c||c|c|c||c|c|c|}\hline
Fields   &    $U(1)_{F_1}$  &   $U(1)_{F_2}$  &Fields &
$U(1)_{F_1}$  &   $U(1)_{F_2}$&Fields   &    $U(1)_{F_1}$  &
$U(1)_{F_2}$\\\hline
$Q^{1}_{L,R}$  &   0, 0    &   3, 2    &$Q^{21}_{L,R}$ &   2, 3    &   29, 29  &$Q^{41}_{L,R}$ & 7, 7    &   18, 17 \\
$Q^{2}_{L,R}$  &   0, 0    &   5, 4    &$Q^{22}_{L,R}$ &   3, 4    &   2, 2    &$Q^{42}_{L,R}$ & 7, 6    &   2, 2    \\
$Q^{3}_{L,R}$  &   0, 0    &   9, 8    &$Q^{23}_{L,R}$ &   3, 4    &   12, 12  &$Q^{43}_{L,R}$ & 7, 6    &   12, 12  \\
$Q^{4}_{L,R}$  &   0, 0    &   11, 10  &$Q^{24}_{L,R}$ &   4, 5    &   29, 29  &$Q^{44}_{L,R}$ & 8, 7    &   3, 3    \\
$Q^{5}_{L,R}$  &   0, 0    &   15, 14  &$Q^{25}_{L,R}$ &   5, 5    &   0, 1    &$Q^{45}_{L,R}$ & 7, 7    &   4, 5    \\
$Q^{6}_{L,R}$  &   0, 0    &   17, 16  &$Q^{26}_{L,R}$ &   7, 6    &   0, 0    &$Q^{46}_{L,R}$ & 7, 7    &   6, 7    \\
$Q^{7}_{L,R}$  &   2, 2    &   1, 2    &$Q^{27}_{L,R}$ &   9, 8    &   0, 0    &$Q^{47}_{L,R}$ & 7, 7    &   8, 9    \\
$Q^{8}_{L,R}$  &   2, 2    &   3, 4    &$Q^{28}_{L,R}$ &   11, 10  &   0, 0    &$Q^{48}_{L,R}$ & 7, 7    &   10, 11  \\
$Q^{9}_{L,R}$  &   2, 2    &   5, 6    &$Q^{29}_{L,R}$ &   13, 12  &   0, 0    &$Q^{49}_{L,R}$ & 10, 9   &   3, 3    \\
$Q^{10}_{L,R}$ &   2, 2    &   7, 8    &$Q^{30}_{L,R}$ &   15, 14  &   0, 0    &$Q^{50}_{L,R}$ & 11, 11  &   4, 4    \\
$Q^{11}_{L,R}$ &   2, 2    &   9, 10   &$Q^{31}_{L,R}$ &   15, 15  &   2, 1    &$Q^{51}_{L,R}$ & 11, 11  &   6, 5    \\
$Q^{12}_{L,R}$ &   2, 2    &   11, 12  &$Q^{32}_{L,R}$ &   15, 15  &   4, 3    &$Q^{52}_{L,R}$ & 13, 12  &   4, 4    \\
$Q^{13}_{L,R}$ &   2, 2    &   13, 14  &$Q^{33}_{L,R}$ &   15, 15  &   6, 5    &$Q^{53}_{L,R}$ & 13, 13  &   6, 5    \\
$Q^{14}_{L,R}$ &   2, 2    &   15, 16  &$Q^{34}_{L,R}$ &   5, 5    &   5, 6    &$Q^{54}_{L,R}$ & 6, 6    &   31, 30  \\
$Q^{15}_{L,R}$ &   2, 2    &   17, 18  &$Q^{35}_{L,R}$ &   5, 5    &   8, 7    &$Q^{55}_{L,R}$ & 7, 7    &   22, 23  \\
$Q^{16}_{L,R}$ &   2, 2    &   19, 20  &$Q^{36}_{L,R}$ &   5, 5    &   14, 15  &$Q^{56}_{L,R}$ & 7, 8    &   24, 24  \\
$Q^{17}_{L,R}$ &   2, 2    &   21, 22  &$Q^{37}_{L,R}$ &   5, 5    &   16, 17  &$Q^{57}_{L,R}$ & 7, 7    &   26, 27  \\
$Q^{18}_{L,R}$ &   2, 2    &   23, 24  &$Q^{38}_{L,R}$ &   5, 5    &   20, 21  &$Q^{58}_{L,R}$ & 8, 9    &   26, 26  \\
$Q^{19}_{L,R}$ &   2, 2    &   25, 26  &$Q^{39}_{L,R}$ &   5, 5    &   22, 23  &$Q^{59}_{L,R}$ & 8, 7    &   29, 29  \\\cline{7-9}
$Q^{20}_{L,R}$ &2, 2    &   27, 28  &$Q^{40}_{L,R}$ &   7, 7    &   16, 15
&\multicolumn{3}{|c|}{Cont. on Table~\ref{tbl:GeneralQ2}}\\\hline
\end{tabular}
\end{center}
\end{table}

\begin{table}
\begin{center}
\caption{Charge assignments for the heavy quark doublets to
be used in a generalized model.  The given values are for the case
where all coefficients are of the form $(H^\dag H/M^2)^n$. See
Tables \ref{tbl:Replacements1}--\ref{tbl:Replacements6} to make the
necessary changes for the different
Lagrangians.}\label{tbl:GeneralQ2}
\begin{tabular}{|c|c|c||c|c|c||c|c|c|}\hline
Fields   &    $U(1)_{F_1}$  &   $U(1)_{F_2}$  &Fields &
$U(1)_{F_1}$  &   $U(1)_{F_2}$&Fields   &    $U(1)_{F_1}$  &
$U(1)_{F_2}$\\\hline \multicolumn{3}{|c||}{Cont. from Table~\ref{tbl:GeneralQ}}&$Q^{67}_{L,R}$ &   9, 9    &   16,15&$Q^{75}_{L,R}$ &   13, 13  &   12, 11\\\cline{1-3}
$Q^{60}_{L,R}$ &   10, 9   &   29, 29  &$Q^{68}_{L,R}$ &   9, 9    &   18, 17 &$Q^{76}_{L,R}$ &   13, 13  &   14, 13\\
$Q^{61}_{L,R}$ &   8, 7    &   33, 33  &$Q^{69}_{L,R}$ &   11, 11  &   12, 11 &$Q^{77}_{L,R}$ &   13, 13  &   18, 17\\
$Q^{62}_{L,R}$ &   10, 9   &   33, 33  &$Q^{70}_{L,R}$ &   11, 11  &   14, 13 &$Q^{78}_{L,R}$ &   13, 13  &   20, 19\\
$Q^{63}_{L,R}$ &   7, 8    &   20, 20  &$Q^{71}_{L,R}$ &   11, 11  &   18, 19 &$Q^{79}_{L,R}$ &   13, 13  &   24, 25\\
$Q^{64}_{L,R}$ &   9, 9    &   20, 21  &$Q^{72}_{L,R}$ &   11, 11  &   20, 21 &$Q^{80}_{L,R}$ &   13, 13  &   26, 27\\
$Q^{65}_{L,R}$ &   9, 9    &   10, 9   &$Q^{73}_{L,R}$ &   11, 11  &   24, 25 &$Q^{81}_{L,R}$ &   13, 13  &   30, 31\\
$Q^{66}_{L,R}$ &   9, 9    &   12, 11  &$Q^{74}_{L,R}$ &   11, 11  &   26,27 &$Q^{82}_{L,R}$ &   13, 13  &   32, 33\\\hline
\end{tabular}
\end{center}
\end{table}

\begin{table}
\begin{center}
\caption{Charge assignments for the heavy quark singlets to
be used in a generalized model.  The given values are for the case
where all coefficients are of the form $(H^\dag H/M^2)^n$. See
Tables \ref{tbl:Replacements1}--\ref{tbl:Replacements6} to make the
necessary changes for the different
Lagrangians.}\label{tbl:GeneralUD}
\begin{tabular}{|c|c|c||c|c|c||c|c|c|}\hline
Fields   &    $U(1)_{F_1}$  &   $U(1)_{F_2}$  &Fields   &
$U(1)_{F_1}$  &   $U(1)_{F_2}$&Fields   &    $U(1)_{F_1}$  &
$U(1)_{F_2}$ \\\hline
$U^{1}_{L,R}$   &   0, 0    &   2,1     &$U^{12}_{L,R}$ &   5, 5    &   17, 18  &$U^{23}_{L,R}$ &   11, 10  &   33, 33  \\
$U^{2}_{L,R}$   &   1, 2    &   1, 1    &$U^{13}_{L,R}$ &   5, 5    &   19, 20  &$U^{24}_{L,R}$ &   13, 12  &   33, 33  \\
$U^{3}_{L,R}$   &   0, 0    &   6, 5    &$U^{14}_{L,R}$ &   5, 5    &   23, 24  &$U^{25}_{L,R}$ &   13, 13  &   7, 6    \\
$U^{4}_{L,R}$   &   0, 0    &   8, 7    &$U^{15}_{L,R}$ &   5, 5    &   25, 26  &$U^{26}_{L,R}$ &   13, 13  &   9, 8    \\
$U^{5}_{L,R}$   &   0, 0    &   12, 11  &$U^{16}_{L,R}$ &   5, 6    &   27, 27  &$U^{27}_{L,R}$ &   13, 13  &   11, 10  \\
$U^{6}_{L,R}$   &   0, 0    &   14, 13  &$U^{17}_{L,R}$ &   6, 6    &   28, 29  &$U^{28}_{L,R}$ &   13, 13  &   27, 28  \\
$U^{7}_{L,R}$   &   5, 5    &   3, 2    &$U^{18}_{L,R}$ &   11, 11  &   7, 6    &$U^{29}_{L,R}$ &   13, 13  &   29, 30  \\
$U^{8}_{L,R}$   &   5, 5    &   5, 4    &$U^{19}_{L,R}$ &   11, 11  &   9, 8    &$U^{30}_{L,R}$ &   13, 13  &   21, 22  \\
$U^{9}_{L,R}$   &   5, 5    &   9, 8    &$U^{20}_{L,R}$ &   11, 11  &11, 10  &$U^{31}_{L,R}$ &   13, 13  &   23, 24  \\\cline{7-9}
$U^{10}_{L,R}$  &   5, 5    &   11, 10  &$U^{21}_{L,R}$ &   11, 11  &   15, 16  \\
$U^{11}_{L,R}$  &   5, 5    &   13, 14  &$U^{22}_{L,R}$ &   11, 11  &17, 18  \\\hline
$D^{1}_{L,R}$   &   5, 5    &   12, 13  &$D^{11}_{L,R}$ &   9, 9    &   21, 22   &$D^{21}_{L,R}$ &   11, 11  &   21, 22   \\
$D^{2}_{L,R}$   &   6, 7    &   13, 13  &$D^{12}_{L,R}$ &   8, 9    &   24, 24   &$D^{22}_{L,R}$ &   11, 11  &   23, 24   \\
$D^{3}_{L,R}$   &   7, 7    &   14, 15  &$D^{13}_{L,R}$ &   9, 9    &   25, 26   &$D^{23}_{L,R}$ &   11, 10  &   29, 29   \\
$D^{4}_{L,R}$   &   6, 6    &   32, 31  &$D^{14}_{L,R}$ &   9, 9    &   13, 12   &$D^{24}_{L,R}$ &   11, 11  &   27, 28   \\
$D^{5}_{L,R}$   &   7, 6    &   33, 33  &$D^{15}_{L,R}$ &   9, 9    &   15, 14   &$D^{25}_{L,R}$ &   13, 13  &   15, 14   \\
$D^{6}_{L,R}$   &   7, 7    &   19, 20  &$D^{16}_{L,R}$ &   9, 9    &   9, 8     &$D^{26}_{L,R}$ &   13, 13  &   17, 16   \\
$D^{7}_{L,R}$   &   7, 6    &   27, 27  &$D^{17}_{L,R}$ &   10, 11  &   8, 8     &$D^{27}_{L,R}$ &   13, 13  &   21, 22   \\
$D^{8}_{L,R}$   &   6, 6    &   28, 29  &$D^{18}_{L,R}$ &   12, 13  &   8,8     &$D^{28}_{L,R}$ &   13, 13  &   23, 34   \\\cline{7-9}
$D^{9}_{L,R}$   &   8, 9    &   18, 18  &$D^{19}_{L,R}$ &   14, 15  &   8, 8     \\
$D^{10}_{L,R}$ &   8, 7    &   22, 22   &$D^{20}_{L,R}$ &   15, 15  &   7,
6     \\\cline{1-6}
\end{tabular}
\end{center}
\end{table}

\begin{table}
\begin{center}
\caption{Replacements made to Tables \ref{tbl:GeneralQ}--
\ref{tbl:GeneralUD} when changing the single power coefficient
$(H^\dag H/M^2)$ to $(S^\dag S/M^2)$. The quantum numbers for the
$U(1)_{F_1}$ and $U(1)_{F_2}$ symmetries do not
change.}\label{tbl:Replacements1}
\begin{tabular}{|c|c|c|c|c|}\cline{1-2}\cline{4-5}
Fields   &   $U(1)_S$   &&   Fields
&$U(1)_S$\\\cline{1-2}\cline{4-5}
$D^{1}_{L,R}$   &0&$\mapsto$&$Q^{83}_{L,R}$    &   1   \\
$D^{2}_{L,R}$   &0&$\mapsto$&$Q^{84}_{L,R}$    &   1   \\
$D^{3}_{L,R}$   &0&$\mapsto$&$Q^{85}_{L,R}$ &   1
\\\cline{1-2}\cline{4-5}
$U^{7}_{L,R}$   &0&$\mapsto$&$Q^{86}_{L,R}$    &   1   \\
$U^{8}_{L,R}$   &0&$\mapsto$&$Q^{87}_{L,R}$ &   1
\\\cline{1-2}\cline{4-5}
$U^{18}_{L,R}$ &0&$\mapsto$&$Q^{88}_{L,R}$    &   1   \\
$U^{19}_{L,R}$ &0&$\mapsto$&$Q^{89}_{L,R}$    &   1   \\
$U^{20}_{L,R}$ &0&$\mapsto$&$Q^{90}_{L,R}$ &   1
\\\cline{1-2}\cline{4-5}
\end{tabular}
\end{center}
\end{table}

\begin{table}
\begin{center}
\caption{Replacements made to Tables \ref{tbl:GeneralQ}--
\ref{tbl:GeneralUD} when changing the second power coefficient
$(H^\dag H/M^2)^2$ to $(H^\dag H/M^2)(S^\dag S/M^2)$.
 The quantum numbers for the
$U(1)_{F_1}$ and $U(1)_{F_2}$ symmetries do not
change.}\label{tbl:Replacements2}
\begin{tabular}{|c|c|c|c|c|}\cline{1-2}\cline{4-5}
Fields   &   $U(1)_S$   &&   Fields
&$U(1)_S$\\\cline{1-2}\cline{4-5}
$U^{25}_{L,R}$ &0&$\mapsto$&$Q^{91}_{L,R}$ &   1   \\
$U^{26}_{L,R}$ &0&$\mapsto$&$Q^{92}_{L,R}$ &   1   \\
$U^{27}_{L,R}$ &0&$\mapsto$&$Q^{93}_{L,R}$ & 1
\\\cline{1-2}\cline{4-5}
$U^{14}_{L,R}$ &0&$\mapsto$&$Q^{94}_{L,R}$ &   1   \\
$U^{15}_{L,R}$ &0&$\mapsto$&$Q^{95}_{L,R}$ &   1   \\
$U^{16}_{L,R}$ &0&$\mapsto$&$Q^{96}_{L,R}$ &   1   \\
$U^{17}_{L,R}$ &0&$\mapsto$&$Q^{97}_{L,R}$ & 1
\\\cline{1-2}\cline{4-5}
$D^{21}_{L,R}$ &0&$\mapsto$&$Q^{98}_{L,R}$ &   1   \\
$D^{22}_{L,R}$ &0&$\mapsto$&$Q^{99}_{L,R}$ & 1
\\\cline{1-2}\cline{4-5}
$Q^{65}_{L,R}$ &0&$\mapsto$&$D^{29}_{L,R}$ &   1   \\
$Q^{66}_{L,R}$ &0&$\mapsto$&$D^{30}_{L,R}$ & 1
\\\cline{1-2}\cline{4-5}
\end{tabular}
\end{center}
\end{table}

\begin{table}
\begin{center}
\caption{Replacements made to Tables \ref{tbl:GeneralQ}--
\ref{tbl:GeneralUD} when changing the second power coefficient
$(H^\dag H/M^2)^2$ to $(S^\dag S/M^2)^2$.  The replacements from
Table \ref{tbl:Replacements2} must also be made with these
replacements.  The quantum numbers for the $U(1)_{F_1}$ and
$U(1)_{F_2}$ symmetries do not change.}\label{tbl:Replacements3}
\begin{tabular}{|c|c|c|c|c|}\cline{1-2}\cline{4-5}
Fields   &   $U(1)_S$   &&   Fields
&$U(1)_S$\\\cline{1-2}\cline{4-5}
$D^{25}_{L,R}$ &0&$\mapsto$&$Q^{100}_{L,R}$  &   1   \\
$D^{26}_{L,R}$ &0&$\mapsto$&$Q^{101}_{L,R}$ &   1
\\\cline{1-2}\cline{4-5}
$U^{12}_{L,R}$ &0&$\mapsto$&$Q^{102}_{L,R}$  &   1   \\
$U^{13}_{L,R}$ &0&$\mapsto$&$Q^{103}_{L,R}$ &   1
\\\cline{1-2}\cline{4-5}
$D^{23}_{L,R}$ &0&$\mapsto$&$Q^{104}_{L,R}$  &   1   \\
$D^{24}_{L,R}$ &0&$\mapsto$&$Q^{105}_{L,R}$ &   1
\\\cline{1-2}\cline{4-5}
$Q^{67}_{L,R}$ &0&$\mapsto$&$D^{31}_{L,R}$    &   1   \\
$Q^{68}_{L,R}$ &0&$\mapsto$&$D^{32}_{L,R}$ &   1
\\\cline{1-2}\cline{4-5}
\end{tabular}
\end{center}
\end{table}

\begin{table}
\begin{center}
\caption{Replacements made to Tables \ref{tbl:GeneralQ}--
\ref{tbl:GeneralUD} when changing the third power coefficient
$(H^\dag H/M^2)^3$ to $(H^\dag H/M^2)^2(S^\dag S/M^2)$.
 The quantum numbers for the
$U(1)_{F_1}$ and $U(1)_{F_2}$ symmetries do not
change.}\label{tbl:Replacements4}
\begin{tabular}{|c|c|c|c|c|}\cline{1-2}\cline{4-5}
Fields   &   $U(1)_S$   &&   Fields
&$U(1)_S$\\\cline{1-2}\cline{4-5}
$U^{1}_{L,R}$   &0&$\mapsto$&$Q^{106}_{L,R}$  &   1   \\
$U^{2}_{L,R}$   &0&$\mapsto$&$Q^{107}_{L,R}$ &   1\\\cline{1-2}\cline{4-5}
$D^{4}_{L,R}$   &0&$\mapsto$&$Q^{108}_{L,R}$  &   1   \\
$D^{5}_{L,R}$   &0&$\mapsto$&$Q^{109}_{L,R}$ &   1\\\cline{1-2}\cline{4-5}
$Q^{57}_{L,R}$ &0&$\mapsto$&$D^{33}_{L,R}$    &   1   \\
$Q^{58}_{L,R}$ &0&$\mapsto$&$D^{34}_{L,R}$ &   1\\\cline{1-2}\cline{4-5}
\end{tabular}
\end{center}
\end{table}

\begin{table}
\begin{center}
\caption{Replacements made to Tables \ref{tbl:GeneralQ}--
\ref{tbl:GeneralUD} when changing the third power coefficient
$(H^\dag H/M^2)^3$ to $(H^\dag H/M^2)(S^\dag S/M^2)^2$. The
replacements from Table \ref{tbl:Replacements4} must also be made
with these replacements.  The quantum numbers for the $U(1)_{F_1}$
and $U(1)_{F_2}$ symmetries do not
change.}\label{tbl:Replacements5}
\begin{tabular}{|c|c|c|c|c|}\cline{1-2}\cline{4-5}
Fields              &$U(1)_S$&            &Fields
&$U(1)_S$\\\cline{1-2}\cline{4-5}
$U^{3}_{L,R}$   &0&$\mapsto$&$Q^{110}_{L,R}$  &   1   \\
$U^{4}_{L,R}$   &0&$\mapsto$&$Q^{111}_{L,R}$ &   1\\\cline{1-2}\cline{4-5}
$U^{23}_{L,R}$ &0&$\mapsto$&$Q^{112}_{L,R}$  &   1   \\
$U^{24}_{L,R}$ &0&$\mapsto$&$Q^{113}_{L,R}$ &   1\\\cline{1-2}\cline{4-5}
$Q^{55}_{L,R}$ &0&$\mapsto$&$D^{35}_{L,R}$    &   1   \\
$Q^{56}_{L,R}$ &0&$\mapsto$&$D^{36}_{L,R}$ &   1\\\cline{1-2}\cline{4-5}
\end{tabular}
\end{center}
\end{table}

\begin{table}
\begin{center}
\caption{Replacements made to Tables \ref{tbl:GeneralQ}--
\ref{tbl:GeneralUD} when changing the third power coefficient
$(H^\dag H/M^2)^3$ to $(S^\dag S/M^2)^3$. The replacements from
Tables \ref{tbl:Replacements4} and \ref{tbl:Replacements5} must also
be made with these replacements.  The quantum numbers for the
$U(1)_{F_1}$ and $U(1)_{F_2}$ symmetries do not
change.}\label{tbl:Replacements6}
\begin{tabular}{|c|c|c|c|c|}\cline{1-2}\cline{4-5}
Fields  &   $U(1)_S$  &               &Fields &$U(1)_S$
\\\cline{1-2}\cline{4-5}
$U^{5}_{L,R}$    &0&$\mapsto$&$Q^{114}_{L,R}$ &   1   \\
$U^{6}_{L,R}$    &0&$\mapsto$&$Q^{115}_{L,R}$ &   1
\\\cline{1-2}\cline{4-5}
$U^{28}_{L,R}$   &0&$\mapsto$&$Q^{116}_{L,R}$ &   1   \\
$U^{29}_{L,R}$   &0&$\mapsto$&$Q^{117}_{L,R}$ &   1
\\\cline{1-2}\cline{4-5}
$Q^{63}_{L,R}$   &0&$\mapsto$&$D^{37}_{L,R}$  &   1   \\
$Q^{64}_{L,R}$   &0&$\mapsto$&$D^{38}_{L,R}$  &   1
\\\cline{1-2}\cline{4-5}
\end{tabular}
\end{center}
\end{table}